\documentclass[
reprint,
superscriptaddress,
amsmath,amssymb,
aps,
prl,
]{revtex4-2}

\DeclareUnicodeCharacter{25CF}{$\bullet$}

\usepackage{graphicx}
\usepackage{xcolor}
\usepackage{pdfpages}
\usepackage{amssymb}
\usepackage{amsmath}
\usepackage{amsbsy}
\usepackage{color}
\usepackage{ulem}
\usepackage{epstopdf}

\def\vec#1{\mbox{\boldmath $#1$}}

\graphicspath{{img/}}

\usepackage{xr}
\makeatletter
\AtBeginDocument{\let\LS@rot\@undefined}
\newcommand*{\addFileDependency}[1]{
  \typeout{(#1)}
  \@addtofilelist{#1}
  \IfFileExists{#1}{}{\typeout{No file #1.}}
}
\makeatother
\newcommand*{\myexternaldocument}[1]{%
    \externaldocument[SM-]{#1}%
    \addFileDependency{#1.tex}%
    \addFileDependency{#1.aux}%
}
\myexternaldocument{si}

\begin{document}
\title{Coarse-graining DNA: Symmetry, non-local elasticity and persistence length}

\author{Yair Augusto Guti\'{e}rrez Fosado}
\thanks{joint first author}
\affiliation{School of Physics and Astronomy, University of Edinburgh, Peter Guthrie Tait Road, Edinburgh, EH9 3FD, UK}
\author{Fabio Landuzzi}
\thanks{joint first author}
\affiliation{Centro CMP3VdA, Istituto Italiano di Tecnologia, via Lavoratori Vittime del Col du Mont 28, 11100, Aosta, Italy }
\author{Takahiro Sakaue}
\thanks{corresponding author, sakaue@phys.aoyama.ac.jp}
\affiliation{Department of Physics and Mathematics, Aoyama Gakuin University 5-10-1 Fuchinobe, Chuo-ku, Sagamihara-shi, Kanagawa 252-5258, Japan.}

\begin{abstract}
    While the behavior of double stranded DNA at mesoscopic scales is fairly well understood, less is known about its relation to the rich mechanical properties in the base-pair scale, which is crucial, for instance, to understand DNA-protein interactions and the nucleosome diffusion mechanism. Here, by employing the rigid base pair model, we connect its microscopic parameters to the persistence length. Combined with all-atom molecular dynamic simulations, our scheme identifies relevant couplings between different degrees of freedom at each coarse-graining step. This allows us to clarify how the scale dependence of the elastic moduli is determined in a systematic way encompassing the role of previously unnoticed off site couplings between deformations with different parity.
\end{abstract}

\maketitle

The mechanical properties of DNA play a vital role in fundamental biological processes~\cite{Garcia_BJ,Aggrawal_C_SB,Nelson_2012}. Due to its hierarchical structure, the behavior of DNA depends on the length scales, so does the model to describe it. At mesoscopic length scales, DNA exhibits an entropic elasticity, for which one can employ a generic flexible polymer (FP) model~\cite{deGennes_book}. 
On the scale of $50-100$ nm, the bending and twisting elastic degrees of freedom become apparent and DNA is described by the worm-like chain (WLC) model~\cite{Bustamante_1994, Marko_1994}. 
However, understanding the mechanical behaviour of DNA at even shorter length scales, relevant to, e.g. DNA-protein interactions, requires a more detailed description that takes into account structural features of DNA double helix and sequence specificity. One of the standard models for that purpose is the rigid base-pair chain (RBP) model, in which a DNA molecule is described as a succession of rigid subunits representing base pairs~\cite{Olson_1998,Maddocks_2009}.

What is the relation between FP, WLC and RBP models? To answer this question, a suitable strategy is to leave the sequence specific effect aside and adopt an average base-pair description. Recall that the mechanical behaviour of DNA at the mesoscopic scale probed by single molecule experiments is arguably one of the most successful topics studied in biophysics to date~\cite{Bustamante1992,Bustamante_1994,Perkins_1995,BUSTAMANTE2000279,Bustamante2003,Lipfert_2010,Lipfert2011,Lipfert2012,Marko1995,Odijk1995,Nelson1998,Nomidis_2017}. In typical experiments, the extension of a relatively long DNA molecule of the order of several tens kbp was measured as a function of an applied force and/or torque~\cite{Bustamante1992,Bustamante_1994,Perkins_1995,BUSTAMANTE2000279,Bustamante2003,Lipfert_2010,Lipfert2011,Lipfert2012}. This provided a rigorous test to existing theories based on FP and WLC, and led to the determination of important physical quantities in DNA elasticity: the bending ($l_{b}$)  and torsional ($l_{t}$)  persistence lengths~\cite{Marko1995,Odijk1995,Nelson1998,Nomidis_2017}. More recently, however, a number of experiments have reported the high flexibility of short DNA fragments characterized by a shorter effective $l_b$, whose origin and relation to the mesoscopic models have been under active debate\cite{Schindler_2018, Yuan_2008, Wiggins_06, Golestanian_2012}.
In this paper, we address this aspect by a combination of theory and all-atom simulations that help us to elucidate the relation between models of DNA elasticity at different scales.

\begin{figure}[t]
	\centering
	\includegraphics[width=0.5\textwidth]{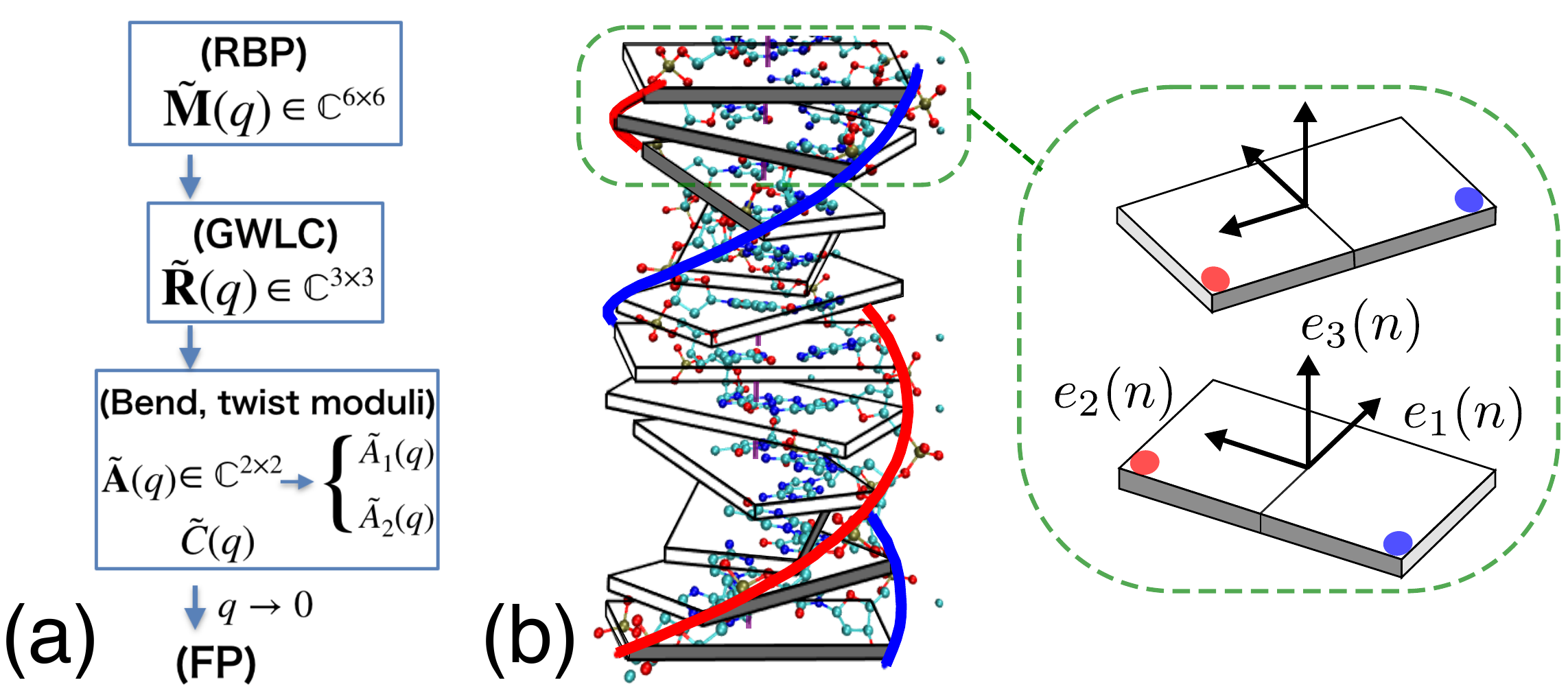}
	\caption{(a)Succession of the coarse graining scheme from RBP model to FP model of DNA through the GWLC model. RBP and GWLC models are characterized by the wave number ($q$) dependent stiffness matrix ${\bf \tilde {M}}(q) \in \mathbb{C}^{6 \times 6}$ and ${\bf \tilde {R}}(q) \in \mathbb{C}^{3 \times 3}$, respectively. From the latter, $q$-dependent bend and twist moduli are extracted, $q \rightarrow 0$ limit of which leads to the persistence lengths characterizing the FP model.
	(b) Definition of the orthonormal frame in RBP model. Starting from the center of the $n$-th brick, ${\vec e}_3(n)$ points to that of $n+1$-th brick, ${\vec e}_1(n)$ lies in the $n$-th brick and points to the major groove direction, which then determine ${\vec e}_2(n) = {\vec e}_3(n) \times {\vec e}_1(n)$.  }
	\label{fig:scheme}
\end{figure}

In principle, a large number of parameters in the RBP model can be evaluated through the analysis of DNA structural fluctuation obtained from all-atom simulations or high resolution crystal structural data. Past studies have often adopted the local free energy assumption to simplify the RBP model and attempted to relate its parameters to persistence lengths~\cite{Lankas_2003, Olson_1998, Becker_2007, Perez_2008}. However, such a locality assumption is not generally correct~\cite{Maddocks_2009,Carlon_2021,Midas_2022}, which implies that the deformation behavior of DNA is length-scale dependent. Here, in order to elaborate on the non-local nature of the RBP model, we present a systematic coarse-graining scheme from RBP to FP through an intermediate model, which we call the generalized worm-like chain (GWLC) model (Fig.~\ref{fig:scheme}). This pinpoints the role of couplings between different degrees of freedom at each coarse-graining step. Our main results are (i) the formulas for the persistence lengths in terms of the RBP model parameters in the long wavelength limit, which are in good agreement with known values, and (ii) a quantification of the scale dependent DNA elasticity, which is compatible with the high flexibility of short DNA fragments~\cite{Schindler_2018, Yuan_2008, Wiggins_06, Golestanian_2012}. Of crucial importance for our paper is the symmetry argument based on the molecular structure of DNA double helix~\cite{Marko_1994}. This symmetry restricts the form of the free energy and hence, it classifies the coupling terms either as symmetric or anti-symmetric. The latter, although largely neglected in literature, produces off-site correlations between deformation degrees of freedom with opposite parity, and affects the high DNA flexibility on the scale of several base-pairs.

\begin{figure*}[]
	\centering
	\includegraphics[width=1.0\textwidth]{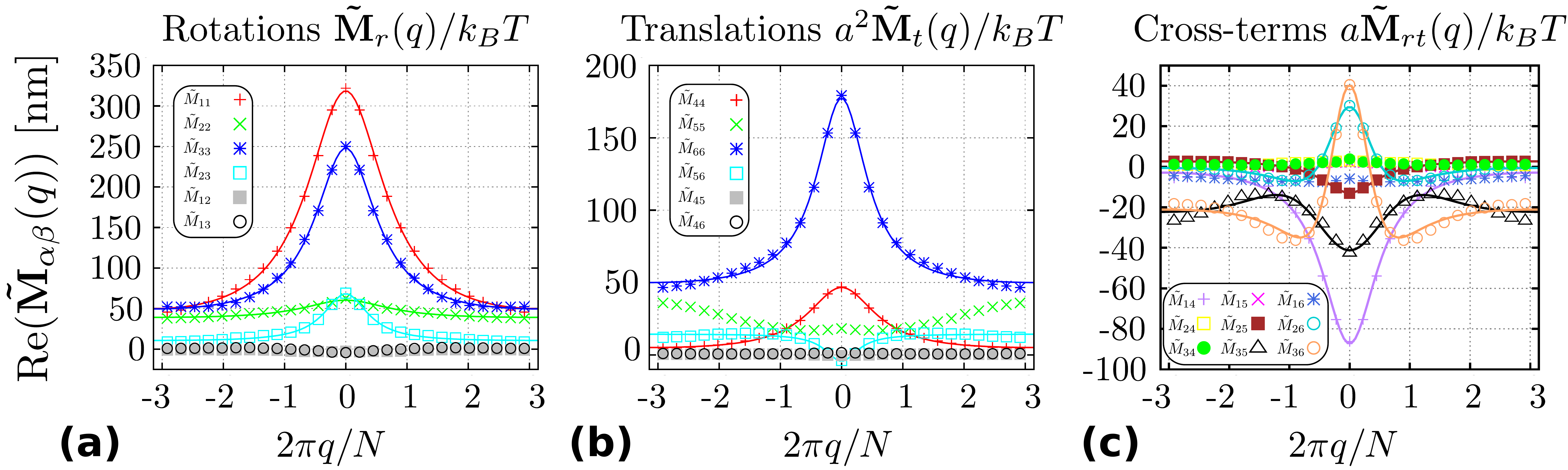}
	\caption{Real part of the stiffness matrix ${\bf {\tilde M}}(q)$ numerically evaluated using Eq.~(\ref{equi-partition}) represented in units of length.	Lines are the fit functions, where we assume the exponential correlations along DNA~\cite{SI}. From left to right, plots show the components of ${\bf {\tilde M}}_r(q)/k_{B}T$, $a^{2}{\bf {\tilde M}}_t(q)/k_{B}T$ and $a{\bf {\tilde M}}_{rt}(q)/k_{B}T$, which correspond to rotational, translational, and coupling submatrices, respectively ($k_BT$ is the thermal energy). Note that some components ($M_{12}, M_{13}, M_{45}, M_{46}, M_{24}, M_{34}, M_{15}, M_{16}$) have negligible real part, which instead exhibit characteristic anti-symmetric profile in their imaginary part~\cite{SI}.
	}
	\label{fig:M}
\end{figure*}

In the RBP model, a DNA molecule is described as a succession of rigid subunits representing base pairs, which are numbered by the index $n \in [-N/2,N/2)$ starting from one end of the DNA chain, where $N$ (assumed to be even) is the total number of base pair steps. The position $\vec{r}(n)$ of the $n$-th base pair and its orientation, described by an embedded orthonormal frame $({\vec e}_1(n), {\vec e}_2(n), {\vec e}_3(n))$, determine the DNA configuration (Fig.~\ref{fig:scheme}).
To describe the local deformation of a DNA molecule, one identifies a vector $\vec{\Omega}(n) \in \mathbb{R}^6$ from the relative position and orientation between consecutive base pairs ($n$ and $n+1$), such that it maps the former to the latter~\cite{Dickerson_1989, Olson_2001}. The first three component ${\vec \Omega}_r(n) \equiv (\Omega_1(n), \Omega_2(n), \Omega_3(n))$ represent the rotational angles per unit length, commonly referred as {\it tilt, roll, twist}. Similarly, the last three components ${\vec \Omega}_t(n) \equiv(\Omega_4(n), \Omega_5(n), \Omega_6(n))$ represent the  translational displacements per unit length, commonly referred as {\it shift, slide, rise}. The free energy takes the quadratic form
\begin{eqnarray}
E(\{\delta \vec{ \Omega}(n) \})=\frac{a}{2}  \sum_{n'} \sum_{n}  \delta \vec{\Omega}^{\rm T}(n') \  {\bf M}(n',n) \ \delta \vec{\Omega}(n),
\label{E_RBP6}
\end{eqnarray}
where sums run over all possible pairs of base-pair steps, $a (=0.34  {\mathrm nm}$) is the average distance between consecutive base pairs, and $ \delta \vec{\Omega}(n) = \vec{\Omega}(n) -  \langle \vec{\Omega}(n) \rangle$ represents the deviation from the thermal average $\langle \vec{\Omega}(n) \rangle$. The stiffness matrix ${\bf M}(n,m) \in \mathbb{R}^{6 \times 6}$ is positive-definite, and describes the interaction between deformations at base pair steps $n$ and $m$. Some remarks on the $M$ matrix are:

(i){\it Reversal and translational invariance for average sequence---} Since we are interested in the average base-pair description, Eq.~(\ref{E_RBP6}) can be rewritten as
\begin{eqnarray}
E(\{\delta \vec{ \Omega}(n) \})
=\frac{a}{2} \sum_{n} \sum_{m} \delta \vec{\Omega}^{\rm T}(n+m) \  {\bf M}(m) \ \delta \vec{\Omega}(n),
\label{E_RBP6_2}
\end{eqnarray}
where $\vec{ \mathrm{M}}$ depends on the separation $m = |n'-n|$ between base pairs~\cite{Carlon_2021}.
Introducing the Fourier transform as 
\begin{eqnarray}
\delta {\tilde {\vec \Omega}}(q) = \sum_{n=-N/2}^{N/2 -1}\delta {\vec \Omega}(n)  \ e^{-2\pi i qn/N},
\end{eqnarray}
with the integer $q \in- [N/2,N/2)$, the energy function (\ref{E_RBP6_2}) is represented as a sum over independent mode contributions
\begin{eqnarray}
E(\{\delta   {\vec  {\tilde \Omega}}(q) \})=\frac{a}{2N} \sum_q \delta {\tilde {\vec\Omega}}^{\rm T}(q) \  {\bf {\tilde M}}(q) \ \delta {\tilde {\vec   \Omega}}(-q).
\label{E_RBP6_2_q}
\end{eqnarray} 

(ii){\it Local approximation is not valid---} From the equi-partition theorem, we find
\begin{eqnarray}
\langle \delta {\tilde {\vec \Omega}}(q)   \delta {\tilde {\vec \Omega}}^{{\mathrm T}}(-q) \rangle = \frac{N k_BT}{a}  {\bf  {\tilde M}}^{-1}(q).
\label{equi-partition}
\end{eqnarray}
If one neglects correlations between deformations at different base pairs, any component of $\langle \delta {\tilde {\vec \Omega}}(q)   \delta {\tilde {\vec \Omega}}^{{\mathrm T}}(-q) \rangle$ just exhibits a white power spectrum, hence the stiffness matrix takes a local form $\vec{ \mathrm{M}}(m) = \vec{ \mathrm{M}}^{o} \delta_{m0}$, making Eq.~(\ref{E_RBP6_2}) to a simple elastic free energy.  However, our numerical simulations (Fig.~\ref{fig:M}) show a colored power spectrum, hence, distal correlations exist in the RBP model of DNA~\cite{Carlon_2021,Midas_2022}. The characteristic bell-shape spectra can be well fitted by a Lorentzian, indicating an exponentially decaying memory along DNA~\cite{SI}, which leads to the softer mechanical behavior in short length scales. The slide degree of freedom ($M_{55}$) is against of this trend exhibiting nonmonotonic q-dependence.
Note also that among the translational degrees of freedom the rise ($M_{66}$) is very stiff, while the shift ($M_{44}$) and the slide are soft. We shall see below that these features make the coupling of the rotational degrees of freedom with the shift and the slide important to determine the mechanical parameters in GWLC.

(iii){\it Conditions imposed by symmetry argument---} Under the reversal of contour coordinate $n \rightarrow {\hat n} \equiv -n$, the deformation vector is shown to be transformed as $ \Omega_i(n) \rightarrow {\hat \Omega}_i({\hat n}) =  \epsilon_i  \Omega_i(n)$, where $\epsilon_1=\epsilon_4=-1$ and $\epsilon_2=\epsilon_3=\epsilon_5 = \epsilon_6=1$, i.e., tilt and shift are odd, and other deformations are even under the reversal operation~\cite{Maddocks_2009, Carlon_2021, SI}. Since the free energy is invariant under this operation, it follows $M_{\alpha \beta}(m) = \epsilon_{\alpha} \epsilon_{\beta} M_{\alpha \beta}(-m) = \epsilon_{\alpha} \epsilon_{\beta} M_{\beta \alpha}(m)$. In Fourier space, the components of the stiffness matrix with the index pair of the same parity, (i.e., $\epsilon_{\alpha} \epsilon_{\beta}=1$), are real ${\tilde M}_{\alpha \beta}(q) = |{\tilde M}_{\alpha \beta}(q)|$, even function of $q$ and constitute the symmetric part of ${\tilde {\mathrm M}}(q)$, i.e., ${\tilde M}_{\alpha \beta}(q) = {\tilde M}_{\beta \alpha}(q) = {\tilde M}_{\alpha \beta}(-q)$. On the other hand, the components with the index pair of the different parity, (i.e., $\epsilon_{\alpha} \epsilon_{\beta}=-1$), are imaginary ${\tilde M}_{\alpha \beta}(q) = i |{\tilde M}_{\alpha \beta}(q)|$, odd function of $q$ and constitute the anti-symmetric part of ${\tilde {\mathrm M}}(q)$, i.e., ${\tilde M}_{\alpha \beta}(q) = -{\tilde M}_{\beta \alpha}(q) =- {\tilde M}_{\alpha \beta}(-q)$. The last condition implies the anti-symmetric components vanish in $q \rightarrow 0$ limit. The results from all-atom simulation are all consistent with this symmetry argument, see Fig.~\ref{fig:M}~\cite{SI}.

(iv){\it Decomposition into rotational and translational components---} In line with the decomposition of the deformation vector ${\vec \Omega}(n)=({\vec \Omega}_r(n), {\vec \Omega}_t(n))$, one can decompose the stiffness matrix ${\bf M}(m)$ as
\begin{eqnarray}
{\bf M}(m) =
\left(
\begin{array}{cc}
{\bf M}_r(m) & {\bf M}_{rt}(m) \\
{\bf M}_{tr}(m) & {\bf M}_t(m)
\end{array}
\right),
\end{eqnarray}
where the submatrices ${\bf M}_r(m), {\bf M}_t(m)  \in \mathbb{R}^{3 \times 3}$ encode the stiffness of RBP model of DNA for the rotational and translational deformations, respectively, and the submatrices ${\bf M}_{rt}(m), {\bf M}_{tr}(m)  \in \mathbb{R}^{3 \times 3}$ represent the rotational-translational coupling with $[M_{tr}(m)]_{\alpha \beta} = \epsilon_{\alpha} \epsilon_{\beta} [M_{rt}(m)]_{\beta \alpha}$. 

In the first step of the coarse-graining, we integrate out the translational degrees of freedom ${\vec \Omega}_t(n)$. Exploiting the Gaussian nature of the energy function~(\ref{E_RBP6_2_q}) one finds the energy function of the GWLC~\cite{Carlon_2021}
\begin{eqnarray}
E_{r}(\{\delta   {\vec  {\tilde \Omega}}_r(q) \})=\frac{a}{2N} \sum_q \delta {\tilde {\vec\Omega}}_r^{\rm T}(q) \  {\bf {\tilde R}}(q) \ \delta {\tilde {\vec   \Omega}}_r(-q),
\label{E_RBP3_q}
\end{eqnarray} 
which depends only on the rotational degrees of freedom ${\vec \Omega}_r (n)$. The corresponding stiffness matrix ${\bf R}(m) \in \mathbb{R}^{3 \times 3}$ is obtained from the original stiffness matrix ${\bf M}(m) \in \mathbb{R}^{6 \times 6}$ as
\begin{eqnarray}
{\bf {\tilde R}}(q) = {\bf {\tilde M}}_r(q) - {\bf {\tilde M}}_{rt}(q)  {\bf {\tilde M}}_t^{-1}(q) {\bf {\tilde M}}_{tr}(q),
\label{R_matrix}
\end{eqnarray}
which is known as the Schur complement of ${\bf {\tilde M}}_t(q)$ in ${\bf {\tilde M}}(q)$~\cite{Schur_complement_book}. 

For conciseness we introduce the following explicit expression relating the components of two matrices.
\begin{eqnarray}
{\tilde R}_{\alpha \beta}(q) = [{\tilde M}_{\alpha \beta}(q)]^{(4,5,6)} \qquad  \alpha,  \beta \in (1,2,3),
\label{R_matrix_component}
\end{eqnarray}
where we introduce the contraction operation
\begin{eqnarray}
[M_{\alpha \beta}]^{(\delta)} \equiv M_{\alpha \beta} - \frac{M_{\alpha \delta} M_{\delta \beta}}{M_{\delta \delta}},
\label{Contraction}
\end{eqnarray}
on the component of an arbitrary matrix $M_{\alpha \beta}$ with the integrated degrees of freedom indicated by the superscript $\delta \  (\neq \alpha, \beta)$.
By applying the contraction operation sequentially, one can readily define the double contraction
\begin{eqnarray}
[ M_{\alpha \beta}]^{(\delta, \gamma)} = [[M_{\alpha \beta}]^{(\delta)}]^{(\gamma)} =  [ [M_{\alpha \beta}]^{(\gamma)}]^{(\delta)},
\label{Multiple_contraction}
\end{eqnarray}
which embodies the effect of integrating two degrees of freedom $\delta, \gamma \ (\neq \alpha, \beta)$ out simultaneously. The order of the contraction steps is irrelevant, and the generalization to the multiple contraction integrating more than two degrees of freedom is straightforward.
 
\begin{figure}[t]
	\centering
	\includegraphics[width=0.4\textwidth]{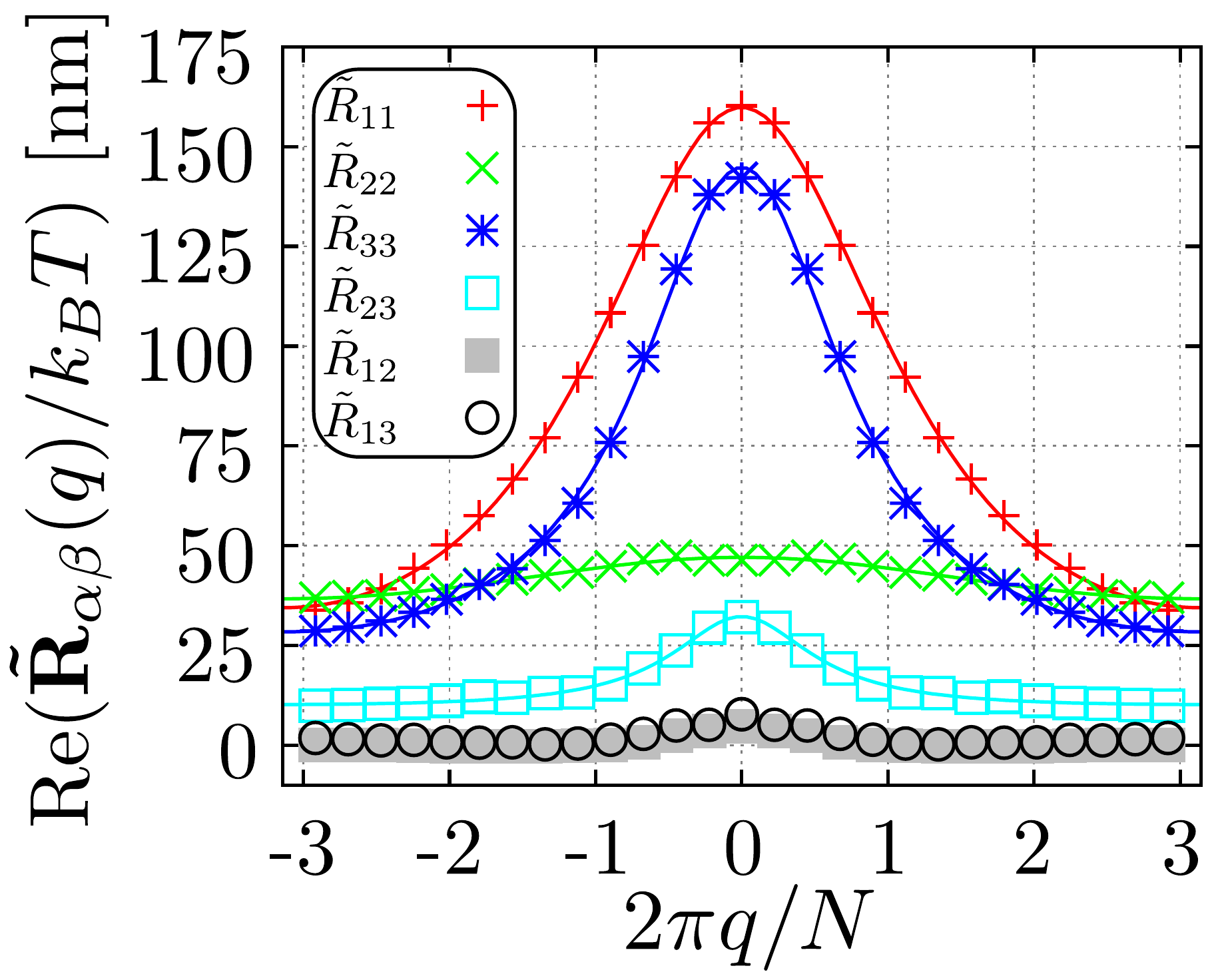}
	\caption{Real part of the stiffness matrix ${\bf {\tilde R}}(q)$ of the GWLC model obtained through contraction operation, Eq.~(\ref{R_matrix}), applied to the ${\bf {\tilde M}}(q)$ matrix. Lines are fit functions with exponential memory along DNA~\cite{SI}. ${\tilde R}_{12}(q)$ and ${\tilde R}_{13}(q)$ have negligible real part in the entire range and exhibit anti-symmetric characteristic profile in their imaginary part~\cite{SI}.
	}
	\label{fig:R}
\end{figure}

The same symmetry argument as for ${\bf M}$ matrix applies to ${\bf R}$ matrix as well~\cite{SI}. The anti-symmetric components $ {\tilde R}_{12}(q),  {\tilde R}_{13}(q)$ are purely imaginary and odd functions of $q$, while all the rest are symmetric and real, even functions of $q$, see Figs.~\ref{fig:R} and Fig.S4~\cite{SI}. While the characteristic $q$ dependence of ${\tilde R}_{\alpha \beta}$ is similar to that of corresponding ${\tilde M}_{r \,  \alpha \beta}$, the magnitude is reduced upon coarse-graining. Indeed, our contraction formula~(\ref{Contraction}) predicts the decrease in the effective rotational stiffness, where the degree of reduction is controlled by the stiffness of the integrated degrees of freedom and the coupling strength with it (the denominator and the numerator in Eq.~(\ref{Contraction}), respectively). 
We observe that while the stiffness of the tilt and the twist decreases roughly by half, that of the roll is rather insensitive. 
This is understood from the negligibly weak roll-slide coupling (${\tilde M}_{25}$) compared to tilt-shift (${\tilde M}_{14}$) and twist-slide (${\tilde M}_{35}$) couplings, whereas the coupling with rise makes only negligible contribution due to its high stiffness (${\tilde M}_{66}$). Importantly, our result on the ${\bf {\tilde R}}(q)$ is in quantitative agreement with the recent report based on the direct analysis of GWLC~\cite{Carlon_2021}, which demonstrates the validity of our coarse-graining procedure.

\begin{figure}[]
	\centering
	\includegraphics[width=0.5\textwidth]{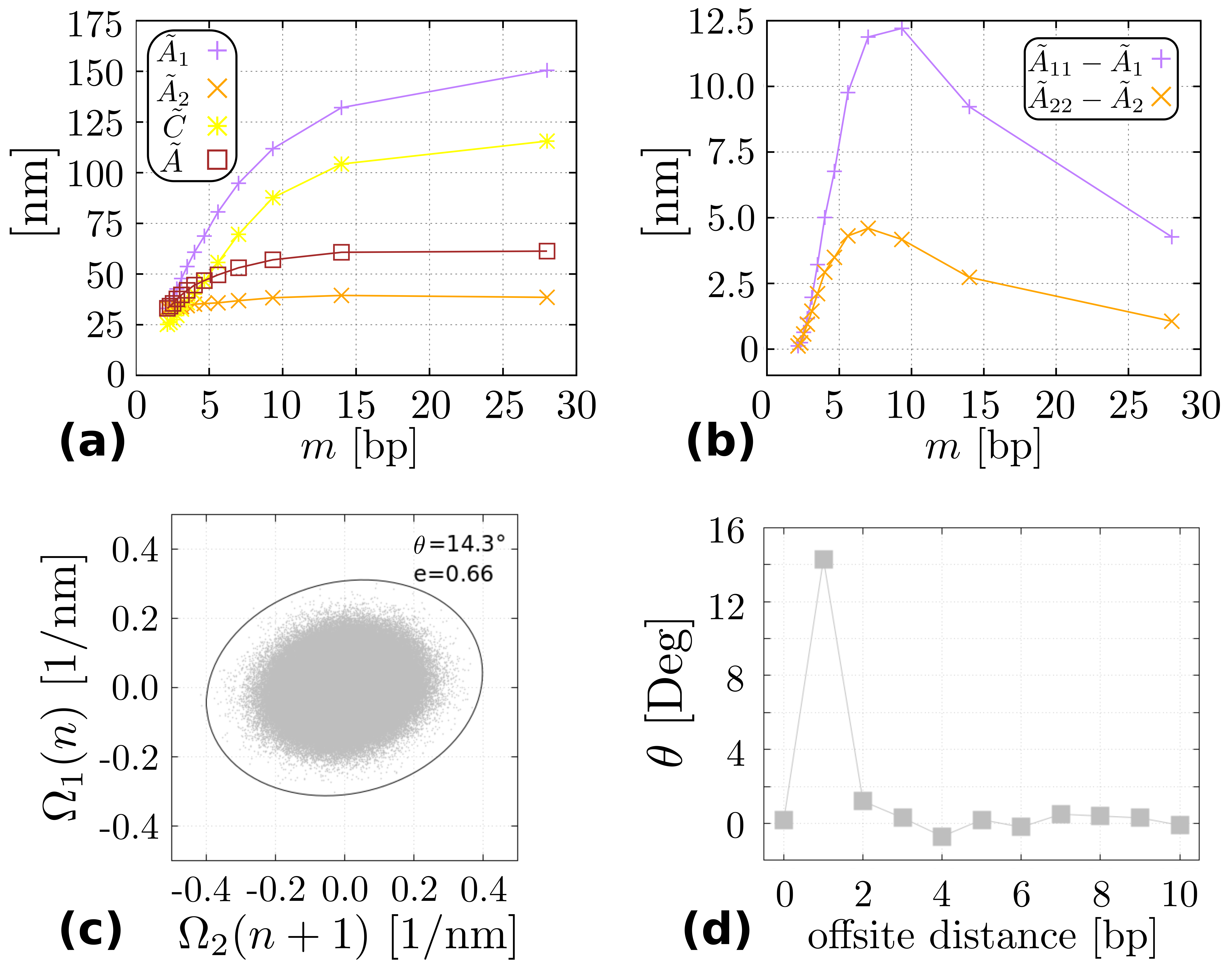}
	\caption{(a) Scale dependent bend (tilt and roll) and twist moduli ${\tilde A}_1(q), {\tilde A}_2(q), {\tilde C}(q)$ as a function of the base-pair separation $m=N/q$. We also show the harmonic mean of tilt and roll ${\tilde A}(q)=2{\tilde A}_1(q){\tilde A}_1(q) /({\tilde A}_1(q) + {\tilde A}_2(q))$. (b) Softening of the tilt ${\tilde A}_{11}(q)-{\tilde A}_{1}(q)$ and the roll ${\tilde A}_{22}(q)-{\tilde A}_{2}(q)$ at finite $q \neq 0$ due to the tilt-roll coupling. (c) Scatter plot of the off-site correlation $\delta \Omega_1(n)$ and $\delta \Omega_2(n+m)$ with $m=1$. The 95 \% confidence ellipse is shown. The tile angle that the ellipse makes with the x axis is $\theta$ and its eccentricity $e$. (d) $\theta$ as function of the off-site distance $m$.}
	\label{fig:AC}
\end{figure}

Next, we integrate out the twisting (or bending) degree of freedom $\Omega_3(n)$ (or $\Omega_1(n)$ and $\Omega_2(n)$). The procedure of this second step of coarse-graining ${\bf {\tilde R}} \rightarrow {\bf {\tilde A}}$ (or ${\bf {\tilde R}} \rightarrow {\tilde C}$) is essentially the same as that of the first step ${\bf {\tilde M}} \rightarrow {\bf {\tilde R}}$~\cite{SI}. Again, the Gaussianity of the energy function~(\ref{E_RBP3_q}) allows us to construct
\begin{eqnarray}
E_{b}(\{\delta   {\vec  {\tilde \Omega}}_b(q) \})=\frac{a}{2N} \sum_q \delta {\tilde {\vec\Omega}}_b^{\rm T}(q) \  {\bf {\tilde A}}(q) \ \delta {\tilde {\vec   \Omega}}_b(-q),
\label{E_RBP_bend_q}
\end{eqnarray} 
which depends only on the bending degrees of freedom ${\vec \Omega}_b (n)= (\Omega_1(n), \Omega_2(n))$, where the bending stiffness matrix ${\bf A}(m) \in \mathbb{R}^{2 \times 2}$ is anti-symmetric with Fourier transform displaying imaginary off-diagonal elements, i.e., ${\tilde A}_{12}(q) = -{\tilde A}_{12}(-q) =  -{\tilde A}_{21}(q)$ as obtained from the rotational stiffness matrix ${\bf R}(m) \in \mathbb{R}^{3 \times 3}$ 
\begin{eqnarray}
{\tilde A}_{\alpha \beta}(q) = [{\tilde R}_{\alpha \beta}(q)]^{(3)}  \qquad  \alpha,  \beta \in (1,2).
\label{A_matrix_component}
\end{eqnarray}
Finally, one can disentangle tilt and roll as ${\tilde A}_{1}(q)=[{\tilde A}_{1 1}(q)]^{(2)}$ and ${\tilde A}_{2}(q)=[{\tilde A}_{2 2}(q)]^{(1)}$ (see Fig.~\ref{fig:AC}). 

Similarly, one can eliminate the bending degrees of freedom and construct the coarse-grained twist energy function
\begin{eqnarray}
E_{t}(\{\delta  {\tilde \Omega}_3(q) \})=\frac{a}{2N} \sum_q \delta {\tilde \Omega}_3^{\rm T}(q) \  {\tilde C}(q) \ \delta {\tilde    \Omega}_3(-q),
\label{E_RBP_twist_q}
\end{eqnarray} 
which depends only on the twisting degrees of freedom $\Omega_3(n)$ with the stiffness $ {\tilde C}(q)$ given by
\begin{eqnarray}
{\tilde C}(q) = [{\tilde R}_{33}(q)]^{(1,2)}.
\label{C_matrix}
\end{eqnarray}

In the long-wavelength limit ($q \rightarrow 0$), all the anti-symmetric components in stiffness matrix vanish~\cite{SI}. Equations~(\ref{A_matrix_component}) and~(\ref{C_matrix}) then lead to
\begin{eqnarray}
{\tilde A}_1(0) = {\tilde R}_{11}(0), \  
{\tilde A}_2(0) =  {\tilde R}_{22}(0) - \frac{{\tilde R}_{23}(0)^2 }{{\tilde R}_{33}(0)},
\label{A_2_eff}
\end{eqnarray}
and the twist modulus
\begin{eqnarray}
{\tilde C}(0) &=& {\tilde R}_{33}(0) -  \frac{{\tilde R}_{23}(0)^2} {{\tilde R}_{22}(0)}.
\label{C_eff}
\end{eqnarray}
These effective moduli evaluated at $q \rightarrow 0$ govern the large scale bending and twisting behaviors of DNA, leading to the persistence lengths
\begin{eqnarray}
l_b = \frac{2}{k_BT}\frac{{\tilde A}_1(0) {\tilde A}_2(0)}{({\tilde A}_1(0) + {\tilde A}_2(0))} \text{ and } \ l_t = \frac{2{\tilde C}(0)}{k_BT}
\label{lb_WLC}
\end{eqnarray}
Equations (\ref{A_2_eff}),~(\ref{C_eff}),~(\ref{lb_WLC}) together with Eq.~(\ref{R_matrix_component}) provide a quantitative connection between RBP model and the WLC, FP models of DNA (see~\cite{SI} for the explicit formula).
From the RBP parameters ${\tilde {\bf M}}(0)$ (Fig.~\ref{fig:M}), our formula leads to ${\tilde A}_1(0)/k_BT = 159$ nm,  ${\tilde A}_2(0)/k_BT = 39$ nm, thus $l_b = 62$ nm, and  ${\tilde C}(0)/k_BT = l_t/2 = 117$ nm, both of which agree with recent reports from coarse-grained simulations~\cite{Mitchell_2017}, all-atom simulations~\cite{Carlon_2021} and  magnetic tweezers experiments~\cite{Lipfert_2010, Nomidis_2017}. Our construction reveals that, in addition to the softened roll and twist due to their coupling~\cite{Carlon_2021,Fosado_2021}, see Eqs.~(\ref{A_2_eff}),~(\ref{C_eff}), a similar softening mechanism is at work in the coarse-graining step from RBP to GWLC model, where the tilt-shift and twist-slide couplings play a decisive role to determine the renormalized stiffness matrix ${\tilde {\bf R}}$, see Eq.~(\ref{R_matrix_component}). 

Several recent experiments have reported that short DNA fragments exhibit much higher flexibility than expected from its ``bulk" elastic property measured in e.g. single molecule experiments~\cite{Schindler_2018, Yuan_2008, Wiggins_06, Golestanian_2012}. In accord with it, the bend and twist moduli as a function of the length scale $m = N/q$ clearly shows the length-scale dependence with softer behavior on small length scale (Fig.~\ref{fig:AC}(a)).
It is worth noting here that the anti-symmetric part of the stiffness matrix is generally non-vanishing at finite $q \neq 0$. Although smaller in magnitude than that of its symmetric counterpart (Fig.S2, S4, S5~\cite{SI}), its effect is non-negligible for the quantitative account for the high flexibility in small scale. As an example, Fig.~\ref{fig:AC}(b) shows the softening of bending response due to the anti-symmetric tilt-roll coupling. This is mirrored in the off-site correlation between the tilt and the roll (Fig.~\ref{fig:AC}(c), (d)). 

In conclusion, we have provided a quantitative connection between models of DNA in different spatial resolutions from the base-pair to mesoscopic scales. We expect that the scale dependent mechanics of DNA is indispensable for the quantitative understanding of DNA-protein interactions, which often induces the tight bend and twist on the scale of several base pair steps. Such features are seen, for instance, in nucleosomes, dictating its structural fluctuation and dynamics~\cite{Skoruppa_PRL}. Also relevant is the DNA response to the intercalators and groove binders, understanding of which is important for better design of the anticancer drugs~\cite{Sahoo_JCP}. Further studies are awaited, which need to address the sequence effect, too.

\begin{acknowledgments}
We thank E. Carlon for stimulating discussions. T.S is supported by JSPS KAKENHI (No. JP18H05529 and JP21H05759) and from MEXT, Japan.
\end{acknowledgments}

\bibliography{bibliography}

\end{document}


\begin{flushleft}
{\bf Supporting Information}
\end{flushleft}


\vspace{2cm}

\section{Symmetry of the stiffness matrix}
\begin{figure}[ht]
    \centering
	\includegraphics[width=0.7\textwidth]{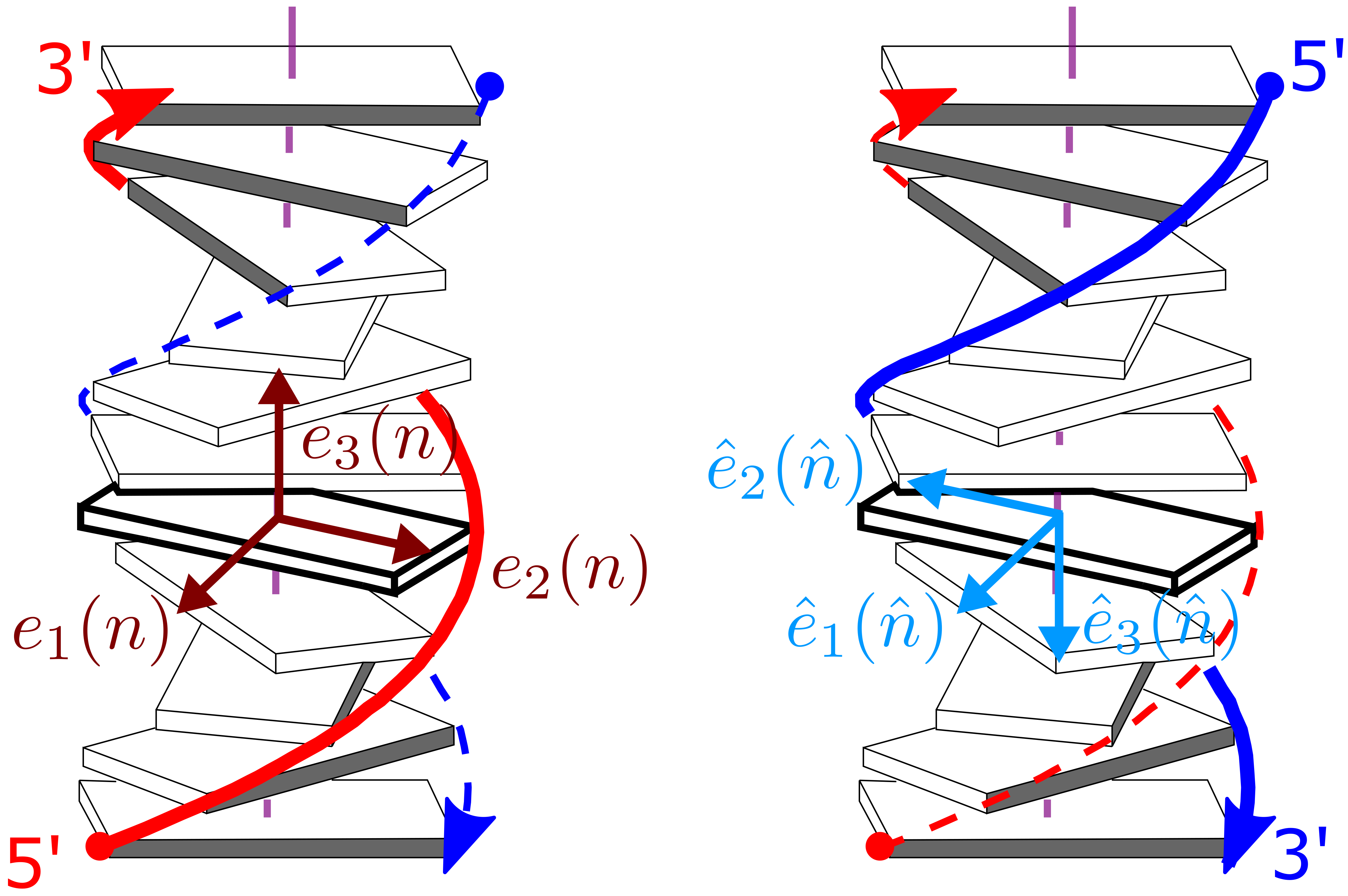}
\caption{Schematic illustrating the relation between original orthonormal frame $({\vec e}_1(n), {\vec e}_2(n), {\vec e}_3(n))$ (left) and that $({\vec {\hat e}}_1({\hat n}), {\vec {\hat e}}_2({\hat n}), {\vec {\hat e}}_3({\hat n}))$ (right) after the reversal of the contour coordinate. The latter is obtained by $\pi$-rotation around ${\vec e}_1$, which is a symmetry operation of the undistorted chain.}
\label{Fig_reversal}
\end{figure}

The double stranded DNA is a helical polymer with a broken reflection symmetry associated to its chiral nature. There are also additional broken symmetries due to the groove asymmetry, i.e., the presence of the major and minor grooves. Thus, as pointed out by Marko and Siggia, the $\pi$-rotation around ${\vec e}_1$, is a symmetry operation of the undistorted chain, but this does not hold for the $\pi$-rotation around ${\vec e}_2$ and ${\vec e}_3$ (Fig.~\ref{Fig_reversal})~\cite{Marko_1994}.
Here we generalize the symmetry argument by Marko and Siggia, and elaborate on the condition for the form of stiffness matrix ${\bf M}(m) \in \mathbb{R}^{6 \times 6}$, see also~\cite{Carlon_2021} for a related discussion. To this end, let us consider the reversal operation $n \rightarrow {\hat n} \equiv -n$ of the contour coordinate as in the main text. To set the contour coordinate, we first take one of the two DNA strands as reference, and number each base pair according to the $5' \rightarrow 3'$ direction of the strand. We denote the position of the $n$-th base pair (represented as a rigid brick in the RBP model) as ${\vec r}(n)$, and fix a right-handed orthonormal frame $({\vec e}_1(n), {\vec e}_2(n), {\vec e}_3(n))$ to its center in a standard way~\cite{Dickerson_1989, Olson_2001}.
By convention, ${\vec e}_1(n)$ and ${\vec e}_2(n)$ lie in the base pair plane with ${\vec e}_1(n)$ pointing along the major groove direction, and ${\vec e}_3(n)$ is normal to the base pair plane pointing to the $3'$ direction of the reference strand.

The reversal of contour coordinate $n$ is achieved by using the complementary DNA strand as the reference strand. Let us number base pairs by using the label ${\hat n}$ according to the $5' \rightarrow 3'$ direction of the new reference strand. Given the anti-parallel orientation of the two strands, it follows ${\hat n} = -n$. Applying the same convention as before, we find the relation between the two right-handed orthonormal frames $({\vec {\hat e}}_1({\hat n}), {\vec {\hat e}}_2({\hat n}), {\vec {\hat e}}_3({\hat n})) = ({\vec e}_1(n), -{\vec e}_2(n), -{\vec e}_3(n))$ (See Fig. \ref{Fig_reversal} for a schematic illustration of the coordinate system).

Then, given a generic deformation, it is easy to check which component of rotation angles or the translational displacements changes the sign upon the reversal of contour coordinate. One finds that the tilt angle (the rotation angle about ${\vec e}_1$) changes the sign, while the roll angle (the rotation angle about ${\vec e}_2$) and the twist angle (the rotation angle about ${\vec e}_3$) do not. Similarly, the shift displacement in the ${\vec e}_1$ direction changes the sign, while the slide and the rise displacements in ${\vec e}_2$ and ${\vec e}_3$ directions, respectively, do not. With our definition of the deformation vector ${\vec {\Omega}}(n) \in \mathbb{R}^6 $, these results can be summarized as 
\begin{eqnarray}
{\hat \Omega}_{\alpha}({\hat n}) = \epsilon_{\alpha} \Omega_{\alpha}(n)
\label{Omega_TF}
\end{eqnarray}
with the parity variables $\epsilon_1 = \epsilon_4=-1$ and $\epsilon_2 = \epsilon_3 = \epsilon_5 = \epsilon_6 =1$.

From Eq. (2) in the main text, the free energy of the deformation takes the form
\begin{eqnarray}
E(\left \{ \delta {\vec \Omega}(n) \right \}) = \frac{a}{2}\sum_n \sum_m \sum_{\alpha=1}^6 \sum_{\beta=1}^6 \delta \Omega_{\alpha} (n+m) M_{\alpha \beta}(m) \delta \Omega_{\beta}(n)
\end{eqnarray}
where $M_{\alpha \beta}(m)$ is a component of the stiffness matrix, which couples the $\beta$-component of the deformation vector to that of the $\alpha$-component located at $m$ base pairs away from the former along the contour coordinate. Note that it follows that $M_{\alpha \beta} (m) = M_{\beta \alpha}(-m)$.

Under the reversal of the contour coordinate, the free energy is rewritten as
\begin{eqnarray}
E(\left \{ \delta {\vec {\hat \Omega}}({\hat n}) \right \}) &=& \frac{a}{2}\sum_{\hat n} \sum_{\hat m} \sum_{\alpha=1}^6 \sum_{\beta=1}^6 \delta {\hat \Omega}_{\alpha} ({\hat n}+{\hat m}) M_{\alpha \beta}({\hat m}) \delta{\hat \Omega}_{\beta}({\hat n}) \nonumber \\
&=& \frac{a}{2}\sum_{n} \sum_{ m} \sum_{\alpha=1}^6 \sum_{\beta=1}^6    \epsilon_{\alpha} \epsilon_{\beta} \delta  \Omega_{\alpha} ( n+m) M_{\alpha \beta}( {\hat m}) \delta \Omega_{\beta}( n)
\end{eqnarray}
where we use Eq.~(\ref{Omega_TF}) in the last equality. Since the energy is invariant under the transformation, we find the condition
\begin{eqnarray}
M_{\alpha \beta}(m) = \epsilon_{\alpha}\epsilon_{\beta}M_{\alpha \beta}(-m) =  \epsilon_{\alpha}\epsilon_{\beta}M_{\beta \alpha}(m)
\label{M_symmetry}
\end{eqnarray}
It follows that the component of the stiffness matrix with the index pair of the same parity, i.e., $\epsilon_{\alpha}\epsilon_{\beta}=1$, has a reversal symmetry and constitute the symmetric part of ${\bf M}(m)$, i.e., $M_{\alpha \beta}(m) = M_{\alpha \beta}(-m) =  M_{\beta \alpha}(m)$, while the component of the stiffness matrix with the index pair of the opposite parity, i.e., $\epsilon_{\alpha}\epsilon_{\beta}=-1$, has an anti-reversal symmetry and constitute the anti-symmetric part of ${\bf M}(m)$, i.e., $M_{\alpha \beta}(m) = -M_{\alpha \beta}(-m) = - M_{\beta \alpha}(m)$. 
After the Fourier transform
\begin{eqnarray}
{\tilde M}_{\alpha \beta}(q) = \sum_{m=-N/2}^{N/2-1} M_{\alpha \beta}(m) e^{-2 \pi qm/N}
\end{eqnarray}
the above condition can be rewritten as
\begin{eqnarray}
{\tilde M}_{\alpha \beta}(q) = \epsilon_{\alpha}\epsilon_{\beta} {\tilde M}_{\alpha \beta}(-q) =  \epsilon_{\alpha}\epsilon_{\beta} {\tilde M}_{\beta \alpha}(q)
\label{M_symmetry_q}
\end{eqnarray}
Therefore, the component of the stiffness matrix with the index pair of the same parity, i.e., $\epsilon_{\alpha}\epsilon_{\beta}=1$ is real and even function of $q$ and constitute the symmetric part of ${\tilde {\bf M}}(q)$, i.e., ${\tilde M}_{\alpha \beta}(q) = |{\tilde M}_{\alpha \beta}(q)|= {\tilde M}_{\alpha \beta}(-q) =  {\tilde M}_{\beta \alpha}(q)$, while the component of the stiffness matrix with the index pair of the opposite parity, i.e., $\epsilon_{\alpha}\epsilon_{\beta}=-1$ is imaginary and odd function of $q$ and constitute the anti-symmetric part of ${\tilde {\bf M}}(q)$, i.e., ${\tilde M}_{\alpha \beta}(q) = i |{\tilde M}_{\alpha \beta}(q)| = -{\tilde M}_{\alpha \beta}(-q) = - {\tilde M}_{\beta \alpha}(q)$. 
As shown in Fig. S2, the analysis of the all-atom MD simulation results demonstrate that the stiffness matrix possesses the above property, thus verifies the validity of the present symmetry argument to impose the form of the free energy.

Inserting $m=0$ in eq.~(\ref{M_symmetry}), we find that the on-site part $M_{\alpha \beta}(0)$ has only symmetric components;
\begin{eqnarray}
{\bf M}(0) = \left(
\begin{array}{cccccc}
M_{11}(0) & 0 & 0 & M_{14}(0) & 0 & 0 \\
0 & M_{22}(0) & M_{23}(0) & 0 & M_{25}(0) & M_{26}(0) \\
0 & M_{23}(0) & M_{33}(0) & 0 & M_{35}(0) & M_{36}(0) \\
M_{14}(0) & 0 & 0 & M_{44}(0) & 0 & 0 \\
0 & M_{25}(0) & M_{35}(0) & 0 & M_{55}(0) & M_{56}(0) \\
0 & M_{26}(0) & M_{36}(0) & 0 & M_{56}(0) & M_{66}(0) 
\end{array}
\right)
\label{M_on-site}
\end{eqnarray}
Similarly, inserting $q=0$ in eq.~(\ref{M_symmetry_q}), we find the long-wavelength limit ${\tilde M}_{\alpha \beta}(0)$ of the stiffness matrix is real and has only symmetric components; written as
\begin{eqnarray}
{\tilde {\bf M}}(0) = \left(
\begin{array}{cccccc}
{\tilde M}_{11}(0) & 0 & 0 & {\tilde M}_{14}(0) & 0 & 0 \\
0 & {\tilde M}_{22}(0) & {\tilde M}_{23}(0) & 0 & {\tilde M}_{25}(0) & {\tilde M}_{26}(0) \\
0 & {\tilde M}_{23}(0) & {\tilde M}_{33}(0) & 0 & {\tilde M}_{35}(0) & {\tilde M}_{36}(0) \\
{\tilde M}_{14}(0) & 0 & 0 & {\tilde M}_{44}(0) & 0 & 0 \\
0 & {\tilde M}_{25}(0) & {\tilde M}_{35}(0) & 0 & {\tilde M}_{55}(0) & {\tilde M}_{56}(0) \\
0 & {\tilde M}_{26}(0) & {\tilde M}_{36}(0) & 0 & {\tilde M}_{56}(0) & {\tilde M}_{66}(0) 
\end{array}
\right)
\label{M_q_0}
\end{eqnarray}
The numerical values of these components evaluated by all-atom MD simulations are listed in Eq.~\ref{Mtilde_fromsim}, where as in Fig. 2 in the main text, the rotational, translational and cross-term components are represented as ${\bf M}_r/k_BT$, $a^2{\bf M}_t/k_BT$ and $a {\bf M}_{rt}k_BT$, respectively, such that they have dimension of length [nm].

\begin{eqnarray}
{\tilde {\bf M}}(0) = \left(
\begin{array}{cccccc}
322       & \ccol{-3} & \ccol{-4} &    -86    & \ccol{2}   & \ccol{-6} \\
\ccol{-3} &    63     &    69     &  \ccol{3} &   -13      & 30 \\
\ccol{-4} &    69     &   250     &  \ccol{4} &   -42      & 41 \\
-86       & \ccol{3}  & \ccol{4}  &     47    & \ccol{0}   & \ccol{2} \\
\ccol{2}  &   -13     &   -42     & \ccol{0}  &     18     & -4 \\
\ccol{-6} &    30     &    41     & \ccol{2}  & -4 & 179
\end{array}
\right)
\label{Mtilde_fromsim}
\end{eqnarray}
Here, the components, which are expected to vanish according to the symmetry analysis, are shaded in blue. Although not zero exactly, these components are very small and negligible compared to the other terms allowed by the symmetry. Note that although the component ${\tilde M}_{56}(0)$ is not forbidden by the symmetry, its value is small. However, it becomes larger in magnitude at $q \neq 0$ and also is an even function of $q$, in line with the symmetry argument.

Essentially the same argument applies to the stiffness matrix ${\bf R}(m) \in \mathbb{R}^{3 \times 3}$ of the GWLC model, which is obtained from the RBP model after the integration of the translational degrees of freedom (see the main text). Again we find the condition
\begin{eqnarray}
R_{\alpha \beta}(m) = \epsilon_{\alpha}\epsilon_{\beta}R_{\alpha \beta}(-m) =  \epsilon_{\alpha}\epsilon_{\beta}R_{\beta \alpha}(m)
\label{R_symmetry}
\end{eqnarray}
\begin{eqnarray}
{\tilde R}_{\alpha \beta}(q) = \epsilon_{\alpha}\epsilon_{\beta} {\tilde R}_{\alpha \beta}(-q) =  \epsilon_{\alpha}\epsilon_{\beta} {\tilde R}_{\beta \alpha}(q)
\label{R_symmetry_q}
\end{eqnarray}
As in the case of the ${\bf M}$ matrix, from eqs.~(\ref{R_symmetry}) and~(\ref{R_symmetry_q}) the anti-symmetric components of the ${\bf R}$ matrix vanish at on-site location and long-wavelength limit;
\begin{eqnarray}
{\bf R}(0) = \left(
\begin{array}{ccc}
R_{11}(0) & 0 & 0  \\
0 & R_{22}(0) & R_{23}(0)  \\
0 & R_{23}(0) & R_{33}(0) 
\end{array}
\right)
\label{R_on-site}
\end{eqnarray}
\begin{eqnarray}
{\tilde {\bf R}}(0) = \left(
\begin{array}{ccc}
{\tilde R}_{11}(0) & 0  & 0 \\
0 & {\tilde R}_{22}(0) & {\tilde R}_{23}(0)  \\
0 & {\tilde R}_{23}(0) & {\tilde R}_{33}(0)  
\end{array}
\right)
\label{R_q_0}
\end{eqnarray}
where ${\tilde {\bf R}}(0)$ is a real matrix. The numerical value of these components obtained through the contraction operation of the ${\bf M}$ matrix are listed in Eq~\ref{Rtilde_fromsim}, where as in Fig. 3 in the main text, they are represented as ${\bf R}/k_BT$, such that they have dimension of length [nm].

\begin{eqnarray}
{\tilde {\bf R}}(0) = \left(
\begin{array}{ccc}
  160    & \ccol{5} & \ccol{8} \\
\ccol{5} &   46     & 32 \\
\ccol{8} &   32     & 142
\end{array}
\right)
\label{Rtilde_fromsim}
\end{eqnarray}
As in the previous case for ${\tilde {\bf M}}(0)$, here the components which are expected to vanish according to the symmetry analysis (shaded in blue) are very small and negligible compared to the other terms allowed by the symmetry.

Note that by discarding all the ${\bf R}(m)$ with $m \neq 0$, i.e., local approximation, the GWLC with ${\bf R}(0)$ given by the form ~(\ref{R_on-site}) reduces to the Marko-Siggia model, where $R_{11}(0)$, $R_{22}(0)$ are two bending moduli, $R_{33}(0)$ is the twist modulus, and $R_{23}(0)$ is the twist-bend coupling coefficient~\cite{Marko_1994}.

\section{Coarse-graining:  $\bf{R}  \rightarrow \bf{A}$ and $C$ }
Since the procedure for the second step of coarse-graining ($\bf{R} \rightarrow \bf{A}$ and $C$) is essentially the same as that of the first step ($\bf{ M} \rightarrow \bf{ R}$), we described it briefly in the main text. Here, we provide the more detailed description.

In accord with the decomposition of the rotational deformation vector ${\vec \Omega}_r(n)=({\vec \Omega}_b(n), \Omega_3(n))$, one can decompose the rotational stiffness matrix ${\bf R}(m) \in \mathbb{R}^{3 \times 3}$ as
\begin{eqnarray}
{\bf R}(m) =
\left(
\begin{array}{cc}
{\bf R}_b(m) & {\bf R}_{bt}(m) \\
{\bf R}_{tb}(m) & {\bf R}_{t}(m)
\end{array}
\right)
\end{eqnarray}
where the submatrix
\begin{eqnarray}
&&{\bf R}_b(m) =
\left(
\begin{array}{cc}
R_{11}(m) & R_{12}(m) \\
-R_{12}(m) & R_{22}(m)
\end{array}
\right)  \in \mathbb{R}^{2 \times 2}
\end{eqnarray}
and $ {\bf R}_{t}(m) = (R_{33}(m))  \in \mathbb{R}^{1 \times 1}$ encode the stiffness for bending and twisting deformations, respectively, and the submatrices
\begin{eqnarray}
&&{\bf R}_{bt}(m) =
\left(
\begin{array}{c}
R_{13}(m)  \\
R_{23}(m) 
\end{array}
\right)   \in \mathbb{R}^{2 \times 1} \\
&&{\bf R}_{tb}(m) =
\left(
\begin{array}{cc}
-R_{13}(m) &   R_{23}(m) 
\end{array}
\right)   \in \mathbb{R}^{1 \times 2}
\end{eqnarray}
represent the twist-bend coupling. 

Let us integrate out the twisting degrees of freedom $\Omega_3(n)$. 
The Gaussianity of the energy function of GWLC
\begin{eqnarray}
E_{r}(\{\delta   {\vec  {\tilde \Omega}}_r(q) \})=\frac{a}{2N} \sum_q \delta {\tilde {\vec\Omega}}_r^{\rm T}(q) \  {\bf {\tilde R}}(q) \ \delta {\tilde {\vec   \Omega}}_r(-q)
\label{E_RBP3_q_SI}
\end{eqnarray} 
allows us to construct
\begin{eqnarray}
E_{b}(\{\delta   {\vec  {\tilde \Omega}}_b(q) \})=\frac{a}{2N} \sum_q \delta {\tilde {\vec\Omega}}_b^{\rm T}(q) \  {\bf {\tilde A}}(q) \ \delta {\tilde {\vec   \Omega}}_b(-q)  
\label{E_RBP_bend_q}
\end{eqnarray} 
which depends only on the bending degrees of freedom ${\vec \Omega}_b (n)= (\Omega_1(n), \Omega_2(n))$, where the bending stiffness matrix ${\bf A}(m) \in \mathbb{R}^{2 \times 2}$ is anti-symmetric with Fourier transform displaying imaginary off-diagonal elements, i.e., ${\tilde A}_{12}(q) = -{\tilde A}_{12}(-q) =  -{\tilde A}_{21}(q)$ as obtained from the rotational stiffness matrix ${\bf R}(m) \in \mathbb{R}^{3 \times 3}$ 
\begin{eqnarray}
{\bf {\tilde A}}(q) = {\bf {\tilde R}}_b(q) - {\bf {\tilde R}}_{bt}(q)  {\bf {\tilde R}}_{t}^{-1}(q) {\bf {\tilde R}}_{tb}(q)
\label{A_matrix}
\end{eqnarray} 
The explicit component expression reads
\begin{eqnarray}
{\tilde A}_{\alpha \beta}(q) = [{\tilde R}_{\alpha \beta}(q)]^{(3)}  \qquad  \alpha,  \beta \in (1,2) 
\label{A_matrix_component}
\end{eqnarray}
Finally, one can disentangle tilt and roll as 
\begin{eqnarray}
{\tilde A}_{1}(q)=[{\tilde A}_{1 1}(q)]^{(2)} \quad \rm{and} \quad  {\tilde A}_{2}(q)=[{\tilde A}_{2 2}(q)]^{(1)}
\label{A_A1_A2}
\end{eqnarray}

Similarly, one can eliminate the bending degrees of freedom and construct the coarse-grained energy function
\begin{eqnarray}
E_{t}(\{\delta  {\tilde \Omega}_3(q) \})=\frac{a}{2N} \sum_q \delta {\tilde \Omega}_3^{\rm T}(q) \  {\tilde C}(q) \ \delta {\tilde    \Omega}_3(-q)
\label{E_RBP_twist_q}
\end{eqnarray} 
which depends only on the twisting degrees of freedom $\Omega_3(n)$ with the stiffness $ {\tilde C}(q)$ given by
\begin{eqnarray}
{\tilde C}(q) 
=  {\bf {\tilde R}}_{t}(q) - {\bf {\tilde R}}_{tb}(q)  {\bf {\tilde R}}_{b}^{-1}(q) {\bf {\tilde R}}_{bt}(q) 
= [{\tilde R}_{33}(q)]^{(1,2)}
\label{C_matrix}
\end{eqnarray}

\section{Elastic moduli in long wave-length limit and persistence lengths}
Here we provide the concrete expression of the elastic moduli ${\tilde A}_1(0), {\tilde A}_2(0), {\tilde C}(0)$ in the long wave-length limit  ($q \rightarrow 0$) in terms of the RBP model parameters.
These moduli are related to the persistence lengths. Specifically, the so-called dynamic bending persistence length $l_{b}$ is a harmonic mean of ${\tilde A}_1(0)$ and ${\tilde A}_2(0)$ and the twist persistence length $l_t$ is twice the ${\tilde C}(0)$ aside from the factor of $k_BT$ (see Eq. (18) in the main text). As shown in Eqs. (16) and (17) in the main text, these elastic moduli are related to the parameters of GWLC model and those of more finer grained RBP model as 
\begin{eqnarray}
{\tilde A}_1(0) &=& {\tilde R}_{11}(0) = [{\tilde M}_{11}(0)]^{(4)} \\
{\tilde A}_2(0) &=& [{\tilde R}_{22}(0)]^{(3)} =  [{\tilde M}_{22}(0)]^{(3,5,6)} \\
{\tilde C}(0) &=& [{\tilde R}_{33}(0)]^{(2)} =  [{\tilde M}_{33}(0)]^{(2,5,6)}
\end{eqnarray}
where the second equalities follow from Eq. (9) in the main text.
Note that $ [{\tilde M}_{11}(0)]^{(4,5,6)} =  [{\tilde M}_{11}(0)]^{(4)}$ and  $ [{\tilde M}_{22}(0)]^{(4,5,6)} =  [{\tilde M}_{33}(0)]^{(5,6)}$, $ [{\tilde M}_{33}(0)]^{(4,5,6)} =  [{\tilde M}_{33}(0)]^{(5,6)}$ due to the structure of the stiffness matrix ${\tilde {\bf M}}(0)$ in the long wave-length limit, see Eq.~(\ref{M_q_0}). Applying the contraction operation defined in Eq. (10) in the main text, we obtain the following expressions;
\begin{eqnarray}
{\tilde A}_1(0) &=& {\tilde M}_{11}(0) - \frac{[{\tilde M}_{14}(0)]^2}{{\tilde M}_{44}(0)} \\
{\tilde A}_2(0) &=& {\tilde M}_{22}(0) 
 -\frac{[{\tilde M}_{25}(0)]^2}{{\tilde M}_{55}(0)} 
- \frac{\left( {\tilde M}_{26}(0)- \frac{{\tilde M}_{25}(0) {\tilde M}_{56}(0)}{{\tilde M}_{55}(0)}\right)^2}{{\tilde M}_{66}(0) - \frac{[{\tilde M}_{56}(0)]^2}{{\tilde M}_{55}(0)}} \nonumber \\
&&- \frac{\left({\tilde M}_{23}(0) - \frac{{\tilde M}_{26}(0) {\tilde M}_{35}(0)}{{\tilde M}_{55}(0)}  -  \frac{\left(  {\tilde M}_{26}(0)- \frac{{\tilde M}_{25}(0){\tilde M}_{56}(0)}{{\tilde M}_{55}(0)} \right) \left(   {\tilde M}_{36}(0)- \frac{{\tilde M}_{35}(0){\tilde M}_{56}(0)}{{\tilde M}_{55}(0)} \right)}{{\tilde M}_{66}(0)- \frac{[{\tilde M}_{56}(0)]^2}{{\tilde M}_{55}(0)}}   \right)^2 }{{\tilde M}_{33}(0)- \frac{[{\tilde M}_{35}(0)]^2}{{\tilde M}_{55}(0)}  - \frac{\left( {\tilde M}_{36}(0)- \frac{{\tilde M}_{35}(0){\tilde M}_{56}(0)}{{\tilde M}_{55}(0)}\right)^2}{{\tilde M}_{66}(0)- \frac{[{\tilde M}_{56}(0)]^2}{{\tilde M}_{55}(0)}}  } \\
{\tilde C}(0) &=& {\tilde M}_{33}(0) 
 -\frac{[{\tilde M}_{35}(0)]^2}{{\tilde M}_{55}(0)} 
- \frac{\left( {\tilde M}_{36}(0)- \frac{{\tilde M}_{35}(0) {\tilde M}_{56}(0)}{{\tilde M}_{55}(0)}\right)^2}{{\tilde M}_{66}(0) - \frac{[{\tilde M}_{56}(0)]^2}{{\tilde M}_{55}(0)}} \nonumber \\
&&- \frac{\left({\tilde M}_{23}(0) - \frac{{\tilde M}_{26}(0) {\tilde M}_{35}(0)}{{\tilde M}_{55}(0)}  -  \frac{\left(  {\tilde M}_{26}(0)- \frac{{\tilde M}_{25}(0){\tilde M}_{56}(0)}{{\tilde M}_{55}(0)} \right) \left(   {\tilde M}_{36}(0)- \frac{{\tilde M}_{35}(0){\tilde M}_{56}(0)}{{\tilde M}_{55}(0)} \right)}{{\tilde M}_{66}(0)- \frac{[{\tilde M}_{56}(0)]^2}{{\tilde M}_{55}(0)}}   \right)^2 }{{\tilde M}_{22}(0)- \frac{[{\tilde M}_{25}(0)]^2}{{\tilde M}_{55}(0)}  - \frac{\left( {\tilde M}_{26}(0)- \frac{{\tilde M}_{25}(0){\tilde M}_{56}(0)}{{\tilde M}_{55}(0)}\right)^2}{{\tilde M}_{66}(0)- \frac{[{\tilde M}_{56}(0)]^2}{{\tilde M}_{55}(0)}}  }
\end{eqnarray}

The numerical values of these elastic moduli obtained from the above formula are listed in Table~\ref{table.fit_AC}. This leads to the dynamic persistence lengths $l_{b} =62 $ nm and $l_t = 240 $ nm. These values are consistent with recent reports from simulations and experiments. For example, Mitchell et al have reported $l_b=58.8$ nm from a Monte Carlo simulation of the coarse-grain rigid-base model of B-form DNA~\cite{Mitchell_2017}, and Skoruppa et al have reported $l_B=61$ nm and $l_t/2 = 125$ nm from all-atom MD simulation~\cite{Carlon_2021}. Lipfert et al~\cite{Lipfert_2010} and Nomidis et al~\cite{Nomidis_2017} have, respectively, reported the twist stiffness $109 \pm 4$ nm and $110 \pm 5$ nm, which is identified as ${\tilde C}(0)/k_BT = l_t/2$ in our notation, from the magnetic tweezers experiment.
We note that since the bending persistence length has the static contribution $l_{b, st}$ in addition to the dynamic contribution $l_{b}$, one usually evaluate the experimentally measured persistence length $l_{b,ex}$ as the harmonic mean of these two contributions; $l_{b,ex} = 2 l_{b}l_{b, st}/(l_{b} + l_{b, st})$~\cite{Trifonov_1988}. It is known $ l_{b, st} \gg  l_{b}$, hence the value of $l_{b,ex}$ gets slightly shorter than $l_{b}$ due to the static contribution.

\section{Functional form of the scale-dependent stiffness matrix}
\begin{figure}[ht]
\centering
	\includegraphics[width=0.98\textwidth]{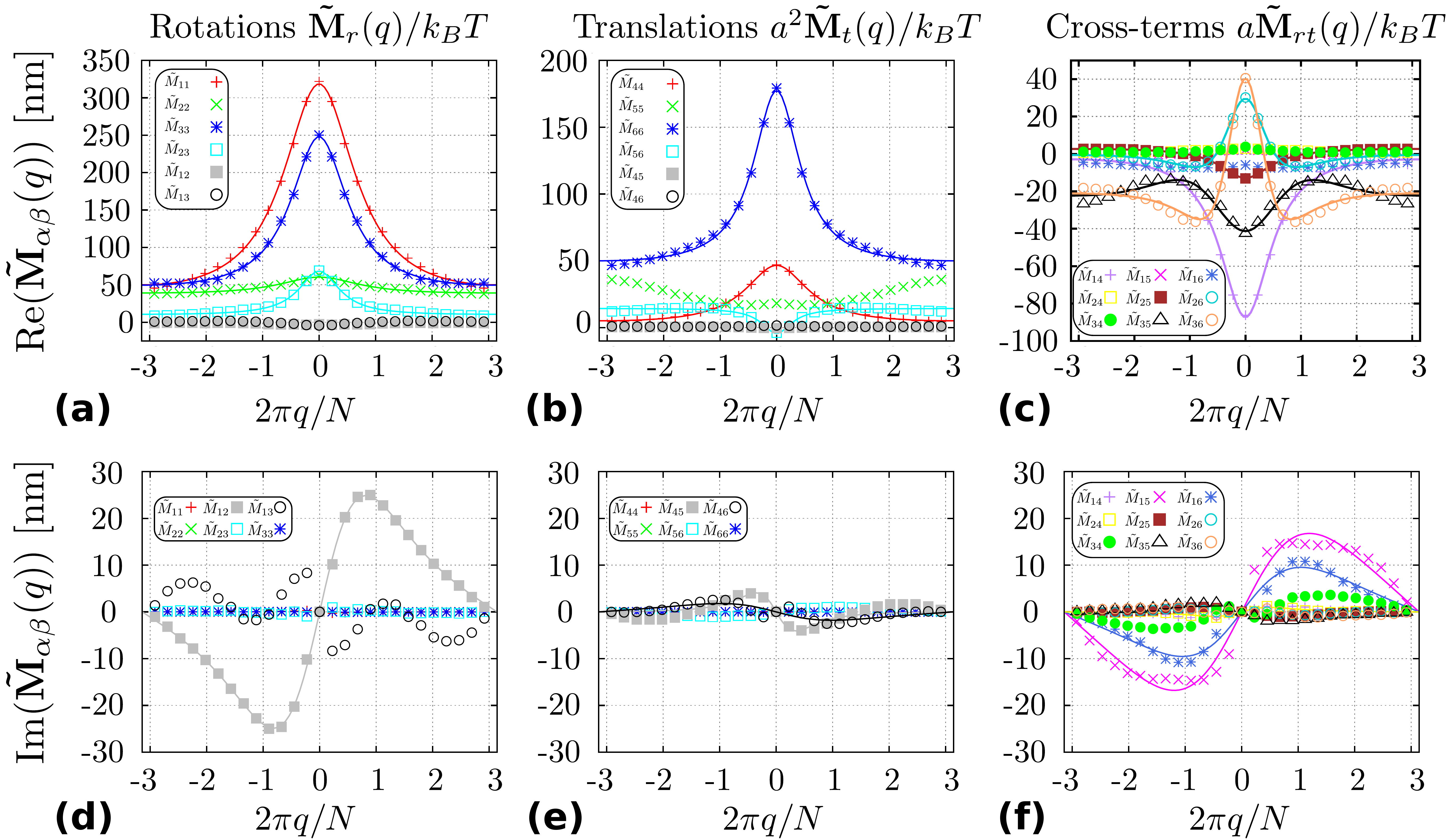}
\caption{Real (top) and imaginary (bottom) parts of the stiffness matrix ${\tilde M}(q)$ of the RBP model numerically evaluated using Eq.(5) in the main text. Points are the numerical data, and lines are fit, whose functional forms are given by Eqs.~(\ref{Eq:ftexp}),~(\ref{Eq:ftodd}) and~(\ref{Eq:ftm2exp}), see the text for the details. To make the quantitative comparison easier, these quantities are represented in units of length. From left to right, plots show the components of ${\bf {\tilde M}}_r(q)/k_BT$, $a^2{\bf {\tilde M}}_t(q)/k_BT$ and $a {\bf {\tilde M}}_{rt}(q)/k_BT$, which correspond to rotational, translational, and coupling submatrices, respectively. Note that the real part plot (top) is the same as Fig. 2 in the main text. The behaviour for $M_{13}$ is rather exceptional, also $M_{55}$ and $M_{34}$ are sensitive to the initial values of the fit probably due to their nonmonotonic behavior close to $q=0$ so we decide to not present the fit.}
\label{Fig_M_im}
\end{figure}

In Fig.~\ref{Fig_M_im}, we show the stiffness matrix ${\tilde {\bf M}}(q)$ of the RBP model, which is computed from the correlations between deformations at different base-pairs observed in all-atom simulations using Eq.(5) in the main text. In agreement with our symmetry analysis, components are classified either into symmetric (even function) or anti-symmetric (odd function) depending on the parity rule for the index pairing. As expected, the real part of the anti-symmetric components is negligibly small. Similarly, the imaginary part of all the symmetric components is again negligible.

We observe that most symmetric components in the scale-dependent stiffness matrix ${\tilde {\bf M}}(q) $ are well fitted by the Lorentzian form
\begin{eqnarray}
{\tilde M}_{\alpha \beta}(q) &=& {\tilde M}_{\alpha \beta}(\infty) + \frac{{\tilde M}_{\alpha \beta}(0)-{\tilde M}_{\alpha \beta}(\infty)}{1 + ({\bar q} \xi_{\alpha \beta})^2} 
\end{eqnarray}
where ${\bar q} \equiv 2 \pi q/Na$ is the wave number, and for convenience, we adopt here the continuum limit ${\tilde A}_1(\infty) \simeq {\tilde A}_1(\pi/a)$ etc.
In real space, this implies the presence of the exponential correlation
\begin{eqnarray}
M_{\alpha \beta}(m) &=& {\tilde M}_{\alpha \beta}(\infty) \delta_{m0} + \frac{({\tilde M}_{\alpha \beta}(0)-{\tilde M}_{\alpha \beta}(\infty))a}{ 2\xi_{\alpha \beta}}  \exp{\left(-\frac{a |m|}{\xi_{\alpha \beta}} \right)}
\label{M_expo}
\end{eqnarray}
In this case, each component of the stiffness matrix $M_{\alpha \beta}(m)$ is characterized by three quantities ${\tilde M}_{\alpha \beta}(0)$, ${\tilde M}_{\alpha \beta}(\infty)$ and $\xi_{M_{\alpha \beta}}$.
This applies to most symmetric components with some exception in the coupling submatrix ${\tilde {\bf M}}_{rt}(q)$, see Fig.\ref{Fig_M_im} (top) and Fig.2 in the main text. The anti-symmetric components have a different functional form, which is odd in $q$, see Fig.\ref{Fig_M_im} (bottom).

In what follows, we attempt the numerical fit for the quantitative determination of each component of the stiffness matrix ${\bf M}(m)$. As an ansatz, let us assume here the following forms:

\begin{subequations}
\begin{eqnarray}
M_{\alpha\beta}(m) &=& D_{\alpha\beta} \delta_{m0} + d_{\alpha\beta} \; e^{\frac{-a\abs{m}}{\xi_{\alpha\beta}}},\label{realMa} \\
M_{\alpha\beta}(m) &=& D_{\alpha\beta} \delta_{m0} + m^{2}d_{\alpha\beta} \; e^{\frac{-a\abs{m}}{\xi_{\alpha\beta}}},\label{realMb} \\
M_{\alpha\beta}(m) &=& md_{\alpha\beta} \; e^{\frac{-a\abs{m}}{\xi_{\alpha\beta}}},\label{realMc}
\end{eqnarray}
\end{subequations}

\noindent The first two equations are associated to the behaviour of the symmetric part of the matrix $M_{\alpha\beta}(m)$; Eq.~(\ref{realMa}), which takes the form of Eq.~(\ref{M_expo}), applies to $M_{11}$,$M_{22}$,$M_{23}$,$M_{33}$,$M_{44}$,$M_{55}$,$M_{56}$,$M_{66}$,$M_{14}$,$M_{25}$ and Eq.~(\ref{realMb}) applies to $M_{26},M_{35},M_{36}$. The latter equation~(\ref{realMc}), represents the anti-symmetric part of the same matrix and is related to the components ($M_{12}, M_{13}, M_{45}, M_{46}, M_{15}, M_{16}, M_{24}, M_{34}$). Therefore, as already stated, each symmetric component of the stiffness matrix is determined by three constant-quantities, $D_{\alpha\beta}$, $d_{\alpha\beta}$ and the correlation length $\xi_{\alpha\beta}$. For anti-symmetric components, which are odd in $q$, $D_{\alpha\beta}=0$ and two constant-quantities, $d_{\alpha\beta}$ and the correlation length $\xi_{\alpha\beta}$ suffice.

At the beginning of this section we mentioned the Lorentzian function, which is known to be the Fourier transform of the continuum exponential function. However, since we deal with discrete data, we compute instead the Discrete Fourier Transform (DFT) of each of the three previous equations. It is worth noting here that we will use the general notation of the DFT,

\begin{equation}
X_{q} = \sum_{m=-l}^{r} x_{m} e^{-2\pi i qm/N}.
\end{equation}

\noindent We recall that this equation transforms a sequence, $\{ x_{m}\}$, of $N$ numbers (real in our case), into a sequence $\{ X_{q} \}$ of complex numbers. In the usual definition of the DFT the limits of the sum are $l=0$ and $r=N-1$. However,since the sequence is $N$-periodic, other sequence of $N$ indices are sometimes used, such as, $l=N/2$ and $r=N/2$.

\subsection{Even terms: Exponential decay}
When the components of the matrix $\mathbf{M}(m)$ follow an exponential decay as the one proposed in Eq.~(\ref{realMa}), the DFT of those terms has the general form:

\begin{eqnarray}
{\tilde M}(q) &=& \sum_{m=-l}^{r} \left[ D\delta_{m0} + d e^{-\frac{a\abs{m}}{\xi}} \right] e^{-2\pi iqm/N} \nonumber \\
        &=& D + d \sum_{m=-l}^{-1} e^{-\frac{a\abs{m}}{\xi} -2\pi iqm/N} + d \sum_{m=0}^{r} e^{-\frac{a\abs{m}}{\xi} -2\pi iqm/N} \nonumber\\
        &=& D + d \sum_{m=1}^{l} e^{-\frac{am}{\xi} + 2\pi iqm/N} + d \sum_{m=0}^{r} e^{-\frac{am}{\xi} -2\pi iqm/N}.
\end{eqnarray}

\noindent where, for simplicity in the notation, we discarded the sub-indices ($\alpha\beta$). Making $m'=m-1$ in the first sum of the above equation we get:

\begin{equation}
{\tilde M}(q) = D + d e^{-\frac{a}{\xi} + 2\pi iq/N} \sum_{m'=0}^{l-1} [e^{-\frac{a}{\xi} + 2\pi iq/N}]^{m'} + d \sum_{m=0}^{r} [e^{-\frac{a}{\xi} -2\pi iq/N}]^{m}.
\end{equation}

\noindent Using the convergence of the geometric series we obtain:

\begin{equation}
{\tilde M}(q) = D + d e^{-\lambda + \varphi} \frac{1-e^{(-\lambda + \varphi) l}}{1-e^{-\lambda + \varphi}} + d \frac{1-e^{(-\lambda - \varphi)(r+1)}}{1-e^{-\lambda - \varphi}},
\end{equation}

\noindent where $\lambda=a/\xi$ and $\varphi=2\pi iq/N$. After adding the previous terms and doing some algebra we get the following result:
\begin{equation}
{\tilde M}(q) = D + d \left[  \frac{\sinh(\lambda)}{\cosh(\lambda)-\cos(2\pi q/N)}  + \frac{e^{-\lambda}\left( e^{-\lambda l}e^{\varphi l} + e^{-\lambda r} e^{-\varphi r} \right) - \left( e^{-\lambda l} e^{\varphi (l+1)} + e^{-\lambda r} e^{-\varphi (r+1)}  \right)   }{\cosh(\lambda)-\cos(2\pi q/N)}\right] 
\end{equation}

\noindent Then, for the interval $l=r=N/2$ we have:

\begin{equation}
{\tilde M}(q) = D + d \left[  \frac{\sinh(\lambda)}{\cosh(\lambda)-\cos(2\pi q/N)}  + \frac{2\cos(\pi q)e^{-\lambda N/2} \left( e^{-\lambda} - \cos(2\pi q/N) \right)}{\cosh(\lambda)-\cos(2\pi q/N)} \right].
\end{equation}

\noindent In the large $N$ limit the previous results reduces to:
\begin{equation}
{\tilde M}(q) = D + d \frac{\sinh(\lambda)}{\cosh(\lambda)-\cos(2\pi q/N)}.
\label{Eq:ftexp}
\end{equation}
We use this equation to fit most of the symmetric components: $M_{11}$,$M_{22}$,$M_{23}$,$M_{33}$,$M_{44}$,$M_{55}$,$M_{56}$,$M_{66}$,$M_{14}$ and $M_{25}$.

\subsection{Odd terms}
The DFT of the $M(m)$ components that behave according to Eq.~(\ref{realMc}), can be obtained in the following way:

\begin{eqnarray}
{\tilde M}(q) &=& d \sum_{m=-l}^{r} m e^{-\frac{a\abs{m}}{\xi}} e^{-2\pi iqm/N} \nonumber \\
        &=&  d \sum_{m=-l}^{-1} me^{-\frac{a\abs{m}}{\xi} -2\pi iqm/N} + d \sum_{m=0}^{r} m e^{-\frac{a\abs{m}}{\xi} -2\pi iqm/N} \nonumber\\
        &=&  d \sum_{m=1}^{l} (-m) e^{-\frac{am}{\xi} + 2\pi iqm/N} + d \sum_{m=0}^{r} m e^{-\frac{am}{\xi} -2\pi iqm/N}.
\end{eqnarray}

\noindent Making $m'=m-1$ in the first sum of the above equation we get:

\begin{equation}
{\tilde M}(q) = - d \sum_{m'=0}^{l-1} (m'+1)e^{(-\lambda + \varphi)(m'+1)} + d \sum_{m=0}^{r} m e^{(-\lambda - \varphi)m}.
\end{equation}

\noindent The terms in the sums can be rewritten in terms of the following derivatives:

\begin{eqnarray}
{\tilde M}(q) &=& d \sum_{m'=0}^{l-1} \frac{\partial}{\partial \lambda} [e^{(-\lambda + \varphi)(m'+1)}] -  d \sum_{m=0}^{r} \frac{\partial}{\partial \lambda} [e^{(-\lambda - \varphi)m}] \nonumber\\
              &=& d \frac{\partial}{\partial \lambda} \left[e^{(-\lambda + \varphi)} \sum_{m'=0}^{l-1} (e^{-\lambda + \varphi})^{m'} - \sum_{m=0}^{r} (e^{-\lambda - \varphi})^{m} \right]
\end{eqnarray}

\noindent Using the criteria of the geometrical series convergence in the above equation:
\begin{eqnarray}
{\tilde M}(q) &=& d \frac{\partial}{\partial \lambda} \left[e^{(-\lambda + \varphi)} \frac{1-e^{(-\lambda + \varphi) l}}{1-e^{-\lambda + \varphi}} - \frac{1-e^{(-\lambda - \varphi)(r+1)}}{1-e^{-\lambda - \varphi}} \right] \nonumber \\
              &=& d \frac{\partial}{\partial \lambda} \left[ \frac{-2\cosh(\lambda)+2e^{\varphi}+e^{(-\lambda+\varphi)l}(e^{-\lambda} - e^{\varphi}) + e^{(-\lambda - \varphi)r}(e^{-\varphi} - e^{-\lambda})}{2\cosh(\lambda)-2\cos(2\pi q/N)} \right]
\end{eqnarray}

\noindent Setting once more the limits of the sum such that $l=r=N/2$:
\begin{eqnarray}
{\tilde M}(q) &=& d \frac{\partial}{\partial \lambda} \left[ \frac{-2\cosh(\lambda)-i 2e^{-\lambda N/2}\sin(\pi +2\pi/N)}{2\cosh(\lambda)-2\cos(2\pi q/N)} \right] \nonumber \\
              &=&- i d \left[ \frac{\sin(2\pi q/N)\sinh(\lambda)}{[\cosh(\lambda)-\cos(2\pi q/N)]^{2}} - e^{-\lambda \frac{N}{2}}\sin(\pi +\frac{2\pi}{N})\frac{\sinh(\lambda)+\frac{N}{2}[\cosh(\lambda)-\cos(2\pi q/N)] }{[\cosh(\lambda)-\cos(2\pi q/N)]^{2}} \right] \nonumber
\end{eqnarray}

\noindent In the long $N$ limit the previous equation becomes

\begin{equation}
{\tilde M}(q) =  - i d \frac{\sin(2\pi q/N)\sinh(\lambda)}{[\cosh(\lambda)-\cos(2\pi q/N)]^{2}}.
\label{Eq:ftodd}
\end{equation}
We use this equation to fit all the anti-symmetric components; $M_{12}$, $M_{13}$, $M_{45}$, $M_{46}$, $M_{15}$, $M_{16}$, $M_{24}$ and $M_{34}$.

\subsection{Even terms: initial growth follow by an exponential decay}
Following a similar process to the one shown for odd terms, here we write the DFT of Eq.~(\ref{realMb}) in terms of the following second derivative:

\begin{eqnarray}
{\tilde M}(q) &=& D +  d \sum_{m=1}^{l} m^{2} e^{-\frac{am}{\xi} + 2\pi iqm/N} + d \sum_{m=0}^{r} m^{2} e^{-\frac{am}{\xi} -2\pi iqm/N}, \nonumber \\
&=& D + d \sum_{m'=0}^{l-1} (m'+1)^{2} e^{(-\lambda + \varphi)(m'+1)} + d \sum_{m=0}^{r} m^{2} e^{(-\lambda -\varphi)m}, \nonumber \\
&=& D + d \frac{\partial^{2}}{\partial \lambda^{2}} \left[e^{(-\lambda + \varphi)} \sum_{m'=0}^{l-1} (e^{-\lambda + \varphi})^{m'} + \sum_{m=0}^{r} (e^{-\lambda - \varphi})^{m} \right], \nonumber \\
&=& D + d \frac{\partial^{2}}{\partial \lambda^{2}} \left[ e^{(-\lambda + \varphi)} \frac{1-e^{(-\lambda + \varphi) l}}{1-e^{-\lambda + \varphi}} + \frac{1-e^{(-\lambda - \varphi)(r+1)}}{1-e^{-\lambda - \varphi}} \right].
\end{eqnarray}

\noindent Then, for the interval $l=r=N/2$ we obtain:
\begin{eqnarray}
{\tilde M}(q) &=& D + d \frac{\partial^{2}}{\partial \lambda^{2}} \left[ \frac{\sinh(\lambda)}{\cosh(\lambda)-\cos(2\pi q/N)}  + \frac{2\cos(\pi q)e^{-\lambda N/2} \left( e^{-\lambda} - \cos(2\pi q/N) \right)}{\cosh(\lambda)-\cos(2\pi q/N)}  \right].
\end{eqnarray}

\noindent After taking the second derivative and the large $N$ limit we get:

\begin{equation}
{\tilde M}(q) = D + d \left[ \frac{2\sinh^{3}(\lambda)}{[\cosh(\lambda)-\cos(2\pi q/N)]^{3}} + \frac{\sinh(\lambda)}{\cosh(\lambda)-\cos(2\pi q/N)} - \frac{3\cosh(\lambda)\sinh(\lambda)}{[\cosh(\lambda)-\cos(2\pi q/N)]^{2}}  \right].
\label{Eq:ftm2exp}
\end{equation}
We use this equation to fit the following symmetric components $M_{26}$, $M_{35}$ and $M_{36}$.

\newpage
\subsection{Fitting to ${\tilde {\bf M}}$ matrix}
\begin{figure}[H]
\centering
	\includegraphics[width=1.0\textwidth]{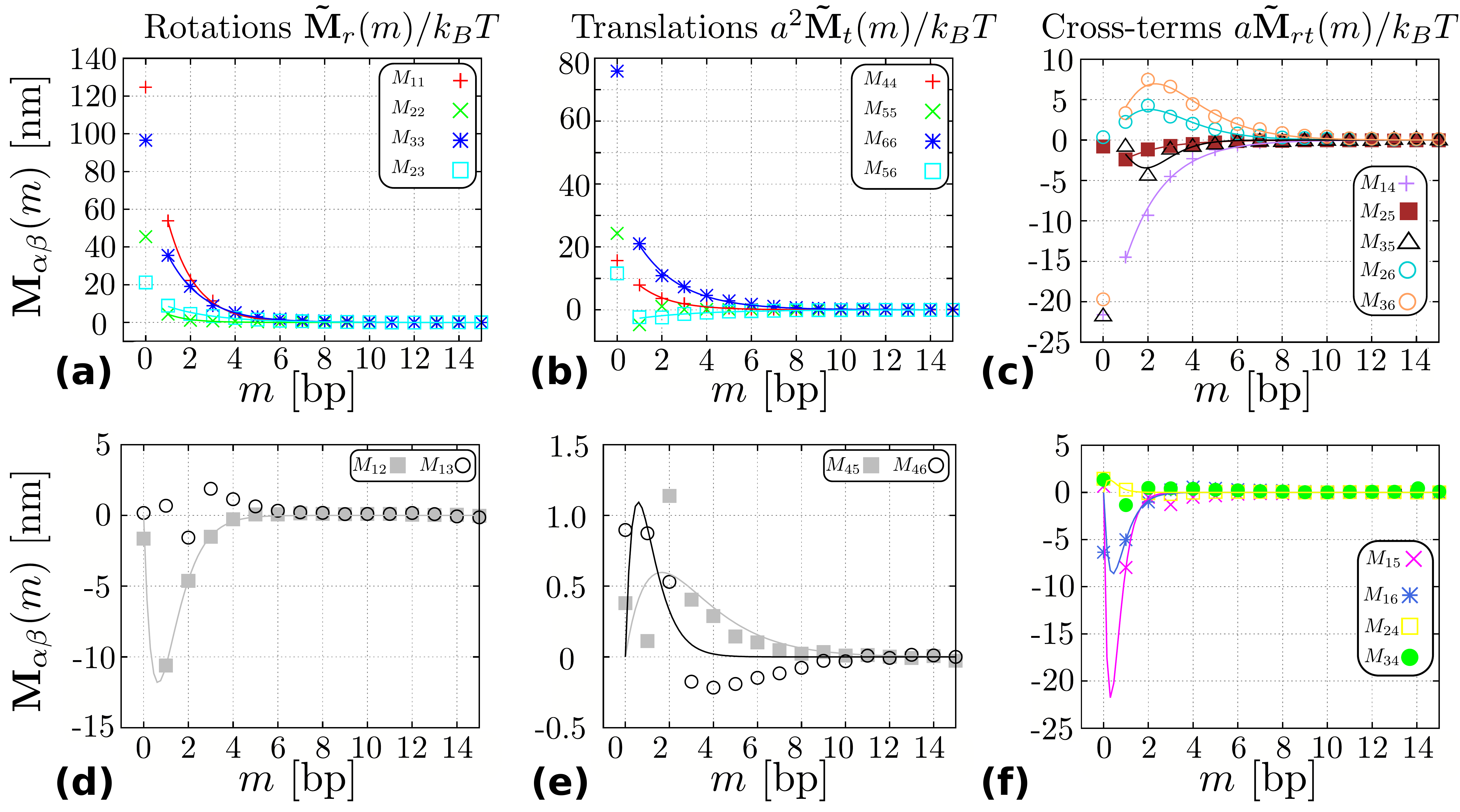}
\caption{Symmetric (top) and antisymmetric (bottom) components of the stiffness matrix ${\bf M}(m)$ as function of the base-pair separation $m$. Points are data obtained from simulations by computing the discrete inverse Fourier transform of the data shown in Fig.~\ref{Fig_M_im}. Lines represent the fit using Eqs.~(\ref{realMa}),~(\ref{realMb}) for symmetric components and Eq.~(\ref{realMc}) for antisymmetric components. The parameters of the fitting ($D_{\alpha \beta}$, $d_{\alpha}$ and $\xi_{\alpha \beta}$) are in agreement with the ones given in Table~\ref{table.fit_M} and obtained by fitting ${\tilde M}_{\alpha\beta}(q)$.}
\label{Fig_M_m}
\end{figure}

In the plots shown in Fig.~\ref{Fig_M_im} for the real and imaginary parts of ${\tilde M}_{\alpha\beta}(q)$, the points represent the data obtained from simulations and lines represent fits using equations (\ref{Eq:ftexp}) (for $M_{11}$, $M_{22}$, $M_{23}$, $M_{33}$, $M_{44}$, [$M_{55}$], $M_{56}$, $M_{66}$, $M_{14}$, $M_{25}$), (\ref{Eq:ftm2exp}) (for $M_{26},M_{35},M_{36}$) and (\ref{Eq:ftodd}) (for $M_{12}$, [$M_{13}$], $M_{45}$, $M_{46}$, $M_{15}$, $M_{16}$, $M_{24}$, [$M_{34}$]). The values obtained from this fit are summarized in Table~\ref{table.fit_M} for the symmetric and anti-symmetric components. The plot of the stiffness matrix ${\bf M}(m)$, which is the inverse Fourier transform of ${\tilde {\bf M}}(q)$, as a function of the base-pair separation $m$ is displayed in Fig.~\ref{Fig_M_m}.
Note that we did not attempt to fit to $M_{55}$, $M_{34}$ and $M_{13}$ since their behaviors are rather exceptional.

\begin{table}[H]
\centering
\resizebox{\textwidth}{!}{%
\begin{tabular}{c|ccccc|ccc|}
\toprule
 & Symmetric       & $D_{\alpha\beta}+d_{\alpha\beta}$ & $d_{\alpha\beta}$ & $\xi_{\alpha\beta}$ & ${\tilde  M_{\alpha\beta}}(q=0)$  &Antisymmetric   & $d_{\alpha\beta}$& $\xi_{\alpha\beta}$  \\
 & ($\alpha\beta$) & [nm]              &  [nm]             &[nm]      &  [nm]       & ($\alpha\beta$) & [nm]             & [nm] \\
\midrule
\midrule
\multirow{4}{*}{Rotations}    & 11 & 122  & 114  & 0.45   &  319  & 12 & 41   & 0.24   \\
                              & 22 &  45  &  10  & 0.40   &  60   & 13 & ---      & ---  \\
                              & 23 &  21  &  13  & 0.79   &  68   &    &          &      \\
                              & 33 &  95  &  65  & 0.56   &  248  &    &          &     \\
\midrule
\multirow{4}{*}{Translations} & 44 &  15 &  15   & 0.51  &   47   & 45 & -1     & 0.57 \\
                              & 55 &  25 & --- & ---     &   15   & 46 & -5     & 0.19 \\
                              & 56 &  12 & -3    & 1.10  &   -4   &    &       &      \\
                              & 66 &  75 &  33   & 0.69  &   177  &    &       &      \\
\midrule
\multirow{5}{*}{Cross-terms}  & 14 & -20.91 & -24.70  & 0.63   &  -87  & 15  & 81   & 0.15 \\
                              & 25 &  -0.67 &  -4.59  & 0.61   &  -13  & 16  & 30   & 0.18 \\
                              & 26 &   5.79 &   5.44  & 0.38   &   30  & 24  & 0.53 & 0.41 \\
                              & 35 & -38.52 & -16.64  & 0.16   &  -41  & 34  &---        & ---   \\
                              & 36 & -11.37 & 8.38    & 0.42   &   40  &     &           &      \\
\bottomrule
\end{tabular}
}
\caption{Elastic parameters of the RBP model obtained from the fitting to all-atom simulations data shown in Fig.~\ref{Fig_M_im}.}
\label{table.fit_M}
\end{table}

\subsection{Fitting to ${\tilde {\bf R}}$ matrix}
\begin{figure}[H]
\centering
	\includegraphics[width=0.82\textwidth]{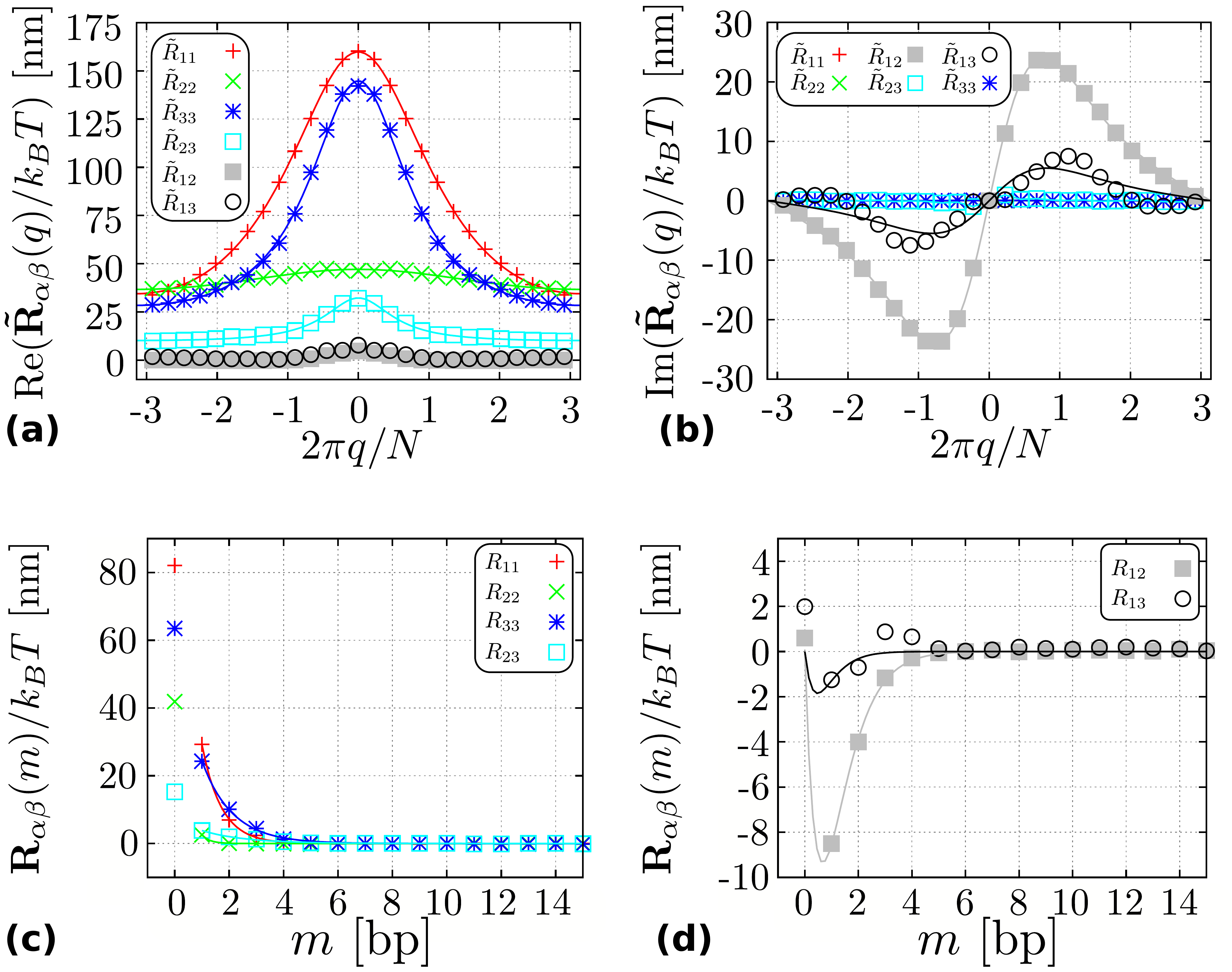}
\caption{Top depicts (a) real and (b) imaginary parts of the stiffness matrix ${\tilde {\bf R}}(q)$ of the GWLC model, which is obtained from ${\tilde {\bf M}}(q)$ through the contraction operation (see Eq. (9) in the main text). Points are the numerical data, and lines are fit whose functional forms are given by Eqs.~(\ref{Eq:ftexp}),~(\ref{Eq:ftodd}), see text for details. From this fit we obtain the values reported in Table~\ref{table.fit_R}. Note that the real part (panel (a)) is the same as Fig. 3 in the main text. Bottom represents components of the ${\bf R}(m)$ matrix as function of the base-pair separation $m$. Data points are obtained from the discrete inverse Fourier transform of the data shown in the top panels. Lines represent the fit using Eqs.~(\ref{realMa}) for symmetric components (panel c) and Eq.~(\ref{realMc}) for antisymmetric components (panel d). The parameters of the fitting ($D_{\alpha \beta}$, $d_{\alpha}$ and $\xi_{\alpha \beta}$) are in agreement with the ones given in Table~\ref{table.fit_R}.}
\label{Fig_R_im}
\end{figure}

The same fitting procedure for the matrix ${\tilde {\bf M}}(q)\in \mathbb{C}^{6 \times 6}$ can be applied to the ${\tilde {\bf R}}(q)\in \mathbb{C}^{3 \times 3}$ matrix, which is obtained upon the coarse-graining by integrating out the translational degrees of freedom. Figure~\ref{Fig_R_im}(top) displays the real and imaginary parts of the stiffness matrix ${\tilde {\bf R}}(q)$. Again, we observe that the data fulfill the requirement from the symmetry argument. In these (top) plots, the points represent the data obtained from simulations and lines represent fits using equations (\ref{Eq:ftexp}) for all the symmetric components ($R_{11}$, $R_{22}$, $R_{23}$, $R_{33}$), and (\ref{Eq:ftodd}) for all the anti-symmetric components ($R_{12}, R_{13}$). The values obtained from this fit are summarized in Table~\ref{table.fit_R}. The plot of the stiffness matrix ${\bf R}(m)$, which is the inverse Fourier transform of ${\tilde {\bf R}}(q)$, as a function of the base-pair separation $m$ is displayed in Fig.~\ref{Fig_R_im}(bottom).

\begin{table}[H]
\centering
\begin{tabular}{|ccccc|ccc|}
\toprule
 Symmetric & $D_{\alpha\beta}+d_{\alpha\beta}$ & $d_{\alpha\beta}$ & $\xi_{\alpha\beta}$ & ${\tilde R_{\alpha\beta}}(q=0)$ &Antisymmetric   & $d_{\alpha\beta}$ & $\xi_{\alpha\beta}$ \\
($\alpha\beta$) &     [nm]     &        [nm]       &       [nm]          &   [nm]  &($\alpha\beta$) &     [nm]        & [nm] \\
\midrule
\midrule
  11 & 80  & 109 &  0.26  & 160  & 12 & 34   & 0.26 \\
  22 & 42  & 62  &  0.11  &  47  & 13 & 9   & 0.24  \\
  23 & 15  & 7   &  0.59  &  32  &    &         & \\
  33 & 62  & 57  &  0.40  &  145 &    &         & \\
\bottomrule
\end{tabular}
\caption{Elastic parameters of GWLC model obtained from fitting to the (top) data in Fig.~\ref{Fig_R_im}.}
\label{table.fit_R}
\end{table}

\subsection{Fitting to ${\tilde {\bf A}}$ matrix and ${\tilde A}_1$, ${\tilde A}_2$, ${\tilde C}$}
\begin{figure}[H]
\centering
	\includegraphics[width=0.84\textwidth]{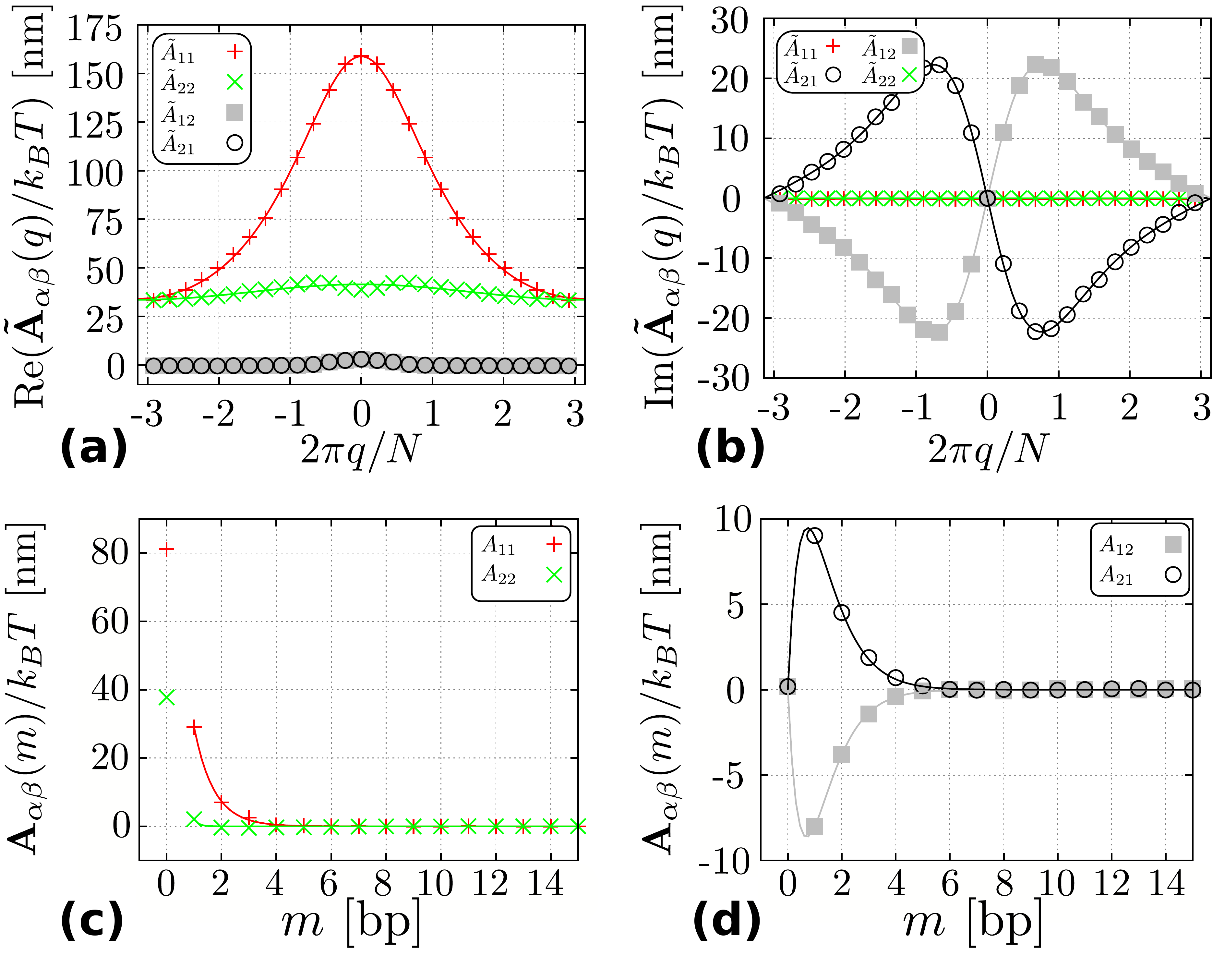}
\caption{Top depicts (a) real and (b) imaginary parts of the bending stiffness matrix ${\tilde {\bf A}}(q)$, which is obtained from ${\tilde {\bf R}}(q)$ through the contraction operation (see Eq. (13) in the main text). Points are the numerical data, and lines are fit whose functional forms are given by Eqs.~(\ref{Eq:ftexp}),~(\ref{Eq:ftodd}), see text for details. From this fit we obtain the values reported in Table~\ref{table.fit_A}. Bottom represents components of the ${\bf A}(m)$ matrix as function of the base-pair separation $m$. Data points are obtained from the discrete inverse Fourier transform of the data shown in the top panels. Lines represent the fit using Eqs.~(\ref{realMa}) for symmetric components (panel c) and Eq.~(\ref{realMc}) for antisymmetric components (panel d). The parameters of the fitting ($D_{\alpha \beta}$, $d_{\alpha \beta}$ and $\xi_{\alpha \beta}$) are in agreement with the ones given in Table~\ref{table.fit_A}.}
\label{Fig_A_im}
\end{figure}

We proceed the fitting for ${\tilde {\bf A}}(q)\in \mathbb{C}^{2 \times 2}$ matrix, and the elastic modulus ${\tilde A}_1(q)$, ${\tilde A}_2(q)$, ${\tilde C}(q)$, which are obtained upon further coarse-graining (see Fig. 1 in the main text). Figure~\ref{Fig_A_im} (top) displays the real and imaginary parts of the bending stiffness matrix ${\tilde {\bf A}}(q)$. In accord with the symmetry argument, ${\tilde A}_{11}(q)$ and ${\tilde A}_{22}(q)$ are real and even function of $q$, while ${\tilde A}_{12}(q) = -{\tilde A}_{21}(q)$ is imaginary and odd function of $q$.
In Figs.~\ref{Fig_A_im}(a-b), the points represent the data obtained from simulations and lines represent fits using equations (\ref{Eq:ftexp}) for $A_{11}$, $A_{22}$, and (\ref{Eq:ftodd}) for $A_{12}$. The values obtained from this fit are summarized in Table~\ref{table.fit_A}. The plot of the stiffness matrix ${\bf A}(m)$, which is the inverse Fourier transform of ${\tilde {\bf A}}(q)$ as a function of the base-pair separation $m$ is displayed in Fig.~\ref{Fig_A_im} (bottom).

\begin{table}[H]
\centering
\begin{tabular}{|ccccc|ccc|}
\toprule
 Symmetric & $D_{\alpha\beta}+d_{\alpha\beta}$ & $d_{\alpha\beta}$ & $\xi_{\alpha\beta}$ & ${\tilde A}_{\alpha\beta}(q=0)$ & Antisymmetric   & $d_{\alpha\beta}$ & $\xi_{\alpha\beta}$ \\
($\alpha\beta$) &     [nm]     &        [nm]       &       [nm]   & [nm]       & ($\alpha\beta$) &     [nm]        & [nm] \\
\midrule
\midrule
  11 & 79  & 106  &  0.27 & 159  & 12 & 31   & 0.26 \\
  22 & 38  & 351  &  0.07 &  41  &    &         &   \\
\bottomrule
\end{tabular}
\caption{Elastic parameters of the bending stiffness matrix obtained from fitting to the simulation data.}
\label{table.fit_A}
\end{table}

\begin{figure}[H]
\centering
	\includegraphics[width=0.98\textwidth]{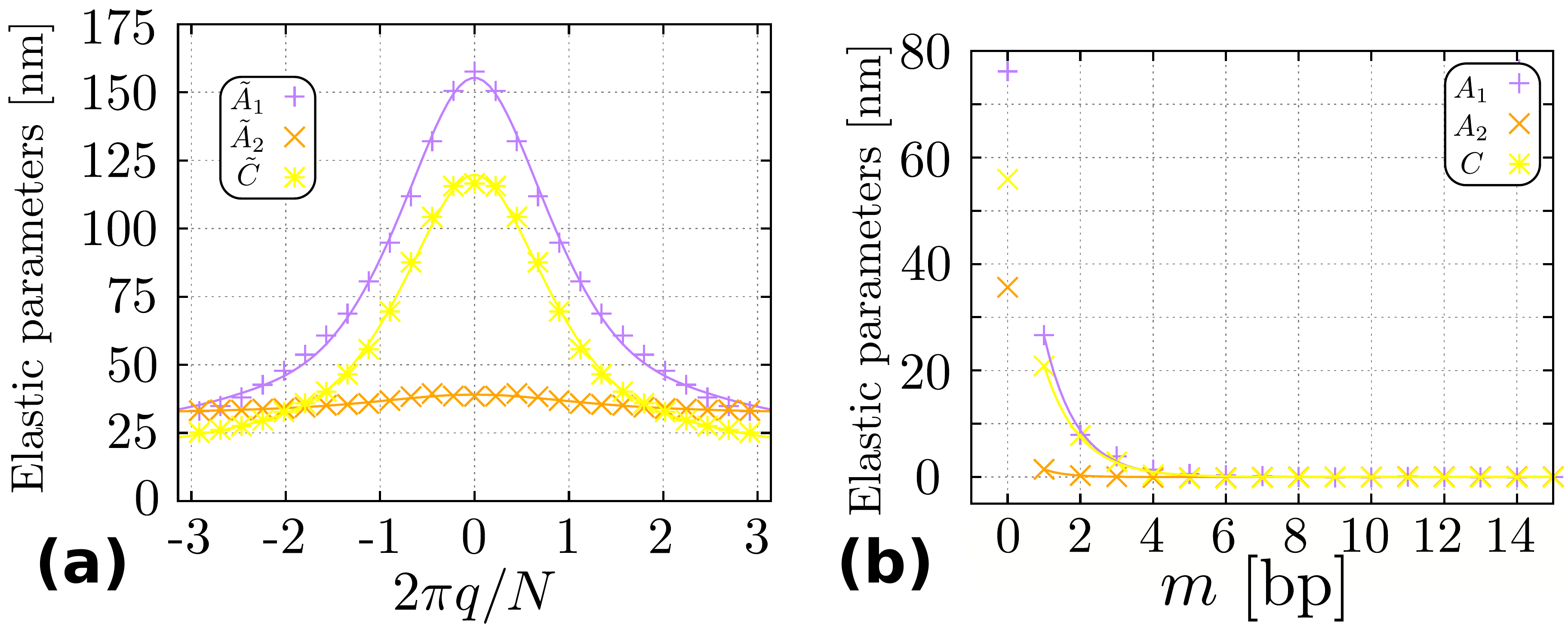}
\caption{(a) Bend moduli ${\tilde A}_1(q)$ and ${\tilde A}_2(q)$ obtained from  ${\tilde {\bf A}}(q)$ matrix through the contraction (see Eq.~(\ref{A_A1_A2})), and the twist modulus ${\tilde C}(q)$ obtained from ${\tilde {\bf R}}(q)$ through the contraction (see Eq.~(\ref{C_matrix}) and Eq. (15) in the main text). Points are the numerical data and lines are fits whose functional forms are given by Eqs.~(\ref{Eq:ftexp}); see the text for the details. From this fit, we obtain the values reported in Table~\ref{table.fit_AC}. Note that this plot is the same as that in Fig. 4(a) in the main text except that the latter is plotted as a function of $m = N/q$. (b) Plots of $A_1(m)$, $A_2(m)$ and $C(m)$ as functions of the base-pair separation $m$. Data points are obtained from the discrete inverse Fourier transform of the data shown in the left panel. Lines represent the fit using Eqs.~(\ref{realMa}). The parameters of the fitting are in agreement with the ones given in Table~\ref{table.fit_AC}.}
\label{Fig_A_C}
\end{figure}

Finally, Fig.~\ref{Fig_A_C}(a) displays the two bend moduli ${\tilde A}_1(q)$, ${\tilde A}_2(q)$ and the twist modulus ${\tilde C}(q)$, which are real quantities. Here, the points represent the data obtained from simulations and lines represent fits using equations (\ref{Eq:ftexp}). The values obtained from this fit are summarized in Table~\ref{table.fit_AC}. The plot of the stiffness $A_1(m)$, $A_2(m)$ and $C(m)$, which are the inverse Fourier transforms of ${\tilde A}_1(q)$, ${\tilde A}_2(q)$ and ${\tilde C}(q)$, respectively, as a function of the base-pair separation $m$ is displayed in Fig.~\ref{Fig_A_C}(b).

\begin{table}[H]
\centering
\begin{tabular}{c|cccc|}
\toprule
  & $D_{\alpha}+d_{\alpha}$ & $d_{\alpha}$ & $\xi_{\alpha}$ & Value at $q=0$ \\
  &     [nm]                          &        [nm]       &       [nm]          &     [nm]                \\
\midrule
\midrule
  $A_{1}$(q) &  75  & 76  & 0.33 & 156 \\
  $A_{2}$(q) &  36  & 7   & 0.21 & 39 \\
      $C$(q) &  55  & 56  & 0.34 & 120 \\
\bottomrule
\end{tabular}
\caption{Elastic parameters of the bend and twist moduli obtained from fitting to the simulation data. Here, the subscript $\alpha$ represents either $A_1$, $A_2$ or $C$.}
\label{table.fit_AC}
\end{table}

\section{Molecular Dynamic Simulation}

Our results are obtained as averages over fragments of B-DNA with the length $32$bps and composed by different sequences of nucleotides: popyA, polyG, polyGC, polyGT, polyGA, polyAT, polyATCG, polyAAATTT, polyAAAATTTT, CGACGTATTACCGTACGATTGGCACTTCACG \fab{\cite{Carlon_2021}}.
The fragments of right handed B-DNA are generated using the code \href{https://casegroup.rutgers.edu/fd_helix.c}{fd helix}
, that uses the same algorithm of the web server \href{http://structure.usc.edu/make-na/ server.html}{make-NA}. 

The software used for the simulation is Gromacs 2020.4\cite{Gromacs} and we adopted the force field Parmbsc1\cite{Orzoco2016}

The systems configurations are obtained solvating the DNA fragments into a triclinic box of TIP3P water molecules, with the addition of $Na^+$ and $Cl^-$ ions such to neutralized the system charge and obtain a $0.15$ Mol salt concentration.

The initial configuration is obtained by applying a standard procedure that consists in consecutive steps: first a minimization to relax the internal stress; followed by a simulation in constant-{NVT}, at a temperature of $300$K, applying constraint to the DNA structure to allow the equilibration of the water molecule and ions in the box. The constant-{NVT} simulation can be divided into two parts, an initial $10.0$ ps where the timestep is $0.5$fs to facilitate the relaxation, and then the timestep is increased to $1.0$fs. Finally the system is equilibrated for $100.0$ps applying a constant-{PNT} at $1.0$atm pressure and $300$K temperature using a timestep of $2.0$fs (this same value is used also to generate the trajectories). All the simulation are carried out in periodic boundaries condition and the modified Berendsen coupling (velocity rescaling with a stochastic term) is used to mimic the effect of a thermostat. After having verified that the system equilibrated to the desired temperature and pressure the final configuration has been used as our initial configuration. This procedure has been applied to each fragment.

The trajectory used to extract the deformation parameters are generated from $200$ns simulation in constant-{NVT} ensemble using $2.0$fs timestep, temperature $300$K. The configuration was dumped each $1.0$ps for a total of 200001 snapshots.
The rotational angles and the translational displacements are extracted from the trajectories using the software Curves+\cite{Curves}. The ${\bf  {\tilde M}}^{-1}$ elements are computed from the Eq.~(5) in the main text using the properties of the Fourier transform. The last two bps from each side of the strand are excluded to avoid the effects due to the fraying (opening of the helix).